\documentclass[12pt,preprint]{aastex}

\shorttitle{K Photoabsorption of O ions}
\shortauthors{Garc\'{\i}a et al.}

\begin{document}

\title{K-shell Photoabsorption of Oxygen Ions}

\author{J. Garc\'{\i}a, C. Mendoza, M. A. Bautista,}
\affil{Centro de F\'{\i}sica, IVIC, Caracas 1020A, Venezuela}
\email{jgarcia@ivic.ve; claudio@ivic.ve; mbautist@ivic.ve}

\author{T. W. Gorczyca,}
\affil{Department of Physics, Western Michigan University, Kalamazoo, MI 49008}
\email{thomas.gorczyca@wmich.edu}

\author{T. R. Kallman,}
\affil{NASA Goddard Space Flight Center, Greenbelt, MD 20771}
\email{timothy.r.kallman@nasa.gov}

\author{and P. Palmeri}
\affil{Astrophysique et Spectroscopie, Universit\'e de Mons-Hainaut, B-7000 Mons, Belgium}
\email{patrick.palmeri@umh.ac.be}

\begin{abstract}
Extensive calculations of the atomic data required for the
spectral modelling of the K-shell photoabsorption of oxygen ions
have been carried out in a multi-code approach. The present level
energies and wavelengths for the highly ionized species (electron
occupancies $2\le N \le 4$) are accurate to within 0.5 eV and 0.02
\AA, respectively. For $N > 4$, lack of measurements, wide
experimental scatter, and discrepancies among theoretical values
are handicaps in reliable accuracy assessments. The radiative and
Auger rates are expected to be accurate to 10\% and 20\%,
respectively, except for transitions involving strongly mixed
levels. Radiative and Auger dampings have been taken into account
in the calculation of photoabsorption cross sections in the
K-threshold region, leading to overlapping lorentzian shaped
resonances of constant widths that cause edge smearing. The
behavior of the improved opacities in this region has been studied
with the {\sc xstar} modelling code using simple constant density
slab models, and is displayed for a range of ionization
parameters.
\end{abstract}

\keywords{atomic processes --- atomic data --- line formation
--- X-rays: general}


\section{Introduction}

The unprecedented  spectral resolution of the grating spectrographs on the
 {\em Chandra} and {\em XMM-Newton} X-ray observatories have unveiled the
diagnostic potential of oxygen K absorption. O~{\sc vii} and
O~{\sc viii} edges are observed  in the spectra of many Seyfert~1
galaxies which have been used by \citet{lee01} to reveal the
existence of a dusty warm absorber in the galaxy MCG -6-30-15.
\citet{ste03} have detected inner-shell transitions of O~{\sc
iii}--O~{\sc vi} in the spectrum of NGC 5548 that point to a warm
absorber that spans three orders of magnitude in ionization
parameter. Moreover, \citet{beh03} have stressed that, in the case
of both Seyfert 1 and Seyfert 2 galaxies, a broad range of oxygen
charge states are usually observed along the line of sight that
must be fitted simultaneously, and may imply strong density
gradients of 2--4 orders of magnitude over short distances.

\citet{sch02} have reported that the spectrum of the archetypical
black hole candidate Cyg X-1 shows a prominent oxygen K edge that
when modelled suggests a deficient O abundance approximately
consistent with an abundance distribution in the interstellar
medium (ISM). Observations of the O~{\sc i} edge and the
equivalent widths of O~{\sc ii}--O~{\sc viii} in the spectrum of
the black-hole binary LMC X-3 by \citet{pag03} lead to upper
limits of the neutral and ionized column densities that rule out a
line-driven stellar wind, and thus suggest a black body with a
multi-temperature disk. A gravitationally redshifted O~{\sc viii}
Ly$\alpha$ observed in absorption in the spectrum of the bursting
neutron star EXO 0748-676 shows, unlike Fe ions, multiple
components consistent with a Zeeman splitting in a magnetic field
of around 1--2$\times 10^9$ G \citep{loe03}. \citet{sak03} has
shown that in optically thick conditions an O~{\sc viii}--N~{\sc
vii} Bowen fluorescence mechanism, caused by the wavelength
coincidence of the respective Ly$\alpha$ (2p--1s) and Ly$\zeta$
(7p--1s) doublets, can suppress O emission while enhancing C and N
thus leading to abundance misinterpretations.

The 1s--2p resonant absorption lines from O ions with electron
occupancies $N\leq 3$ are being used with comparable efficiency in
the detection of the warm-hot intergalactic medium which,
according to simulations of structure formation in the Universe,
contains most of the baryons of the present epoch \citep{cag04}. A
spectral comparison of low and high extinction sources can lead to
the separation of the ISM and instrumental components of the K
edge \citep{dev03}, the former closely resembling the edge
structure of neutral oxygen computed by \citet{mcl98} with the
$R$-matrix method \citep{ber87,sea87}. High-resolution
spectroscopy of the interstellar O K edge in X-ray binaries has
been carried out by \citet{jue04}, discounting oxygen features
from dust and molecular components and providing the first
estimates of the O ionization fractions.

Accurate laboratory wavelength measurements for K-shell resonance
lines in O~{\sc vii} have been reported by \citet{eng95} using
laser pulses focused on compressed powder targets and by
\citet{bei03} with an electron beam ion trap; the latter method
has been recently extended to O~{\sc vi} and O~{\sc v}
\citep{sch04}. K-shell photoabsorption of O~{\sc i} has been
measured by \citet{men96} and \citet{sto97}, and of O~{\sc ii} by
\citet{kaw02} where the role of theoretical data---namely level
energies, wavelengths, $gf$-values, and natural widths---in
experimental interpretation has been emphasized. Laboratory
energies for K-vacancy states in O~{\sc i} and O~{\sc ii} have
been obtained by Auger electron spectrometry \citep{cal93, kra94,
cal94}, but worrisome discrepancies with the photoabsorption
measurements have not as yet been settled.

Such experimental activity has certainly encouraged theoretical
verifications which have been carried out with a variety of atomic
structure and close-coupling methods. Radiative and Auger data
related to the satellite lines of H-, He-, and Li-like oxygen have
been computed in an approach based on the $1/Z$ hydrogenic
expansion \citep{vai71, vai78} and with the multiconfiguration
Dirac--Fock method \citep{che85, che86}. A theoretical study of
the K-shell Auger spectrum of O~{\sc i} has been carried out with
the multiconfiguration Hartree--Fock method \citep{sah94} finding
good agreement with the measurements by \citet{cal93}. Energy
positions and total and partial Auger decay rates for the $[{\rm
1s}]{\rm 2p}^4\ ^{4}{\rm P}$ and $^{2}{\rm P}$ terms of O~{\sc ii}
have been determined by \citet{pet94} with the structure code {\sc
superstructure} \citep{eis74} and {\sc photuc} \citep{sar87}, the
latter being based on an implementation of the close-coupling
approximation of scattering theory. The K-shell photoionization
cross section of the ground state of neutral oxygen has been
computed with the $R$-matrix method by \citet{mcl98}, giving a
detailed comparison with the experimental energy positions for the
$[{\rm 1s}]{\rm 2p}^4(^{4}{\rm P})n{\rm p}\ ^3{\rm P}^{\rm o}$ and
$[{\rm 1s}]{\rm 2p}^4(^{2}{\rm P})n{\rm p}\ ^3{\rm P}^{\rm o}$
Rydberg series. More recently, the $R$-matrix method has also been
used to compute the high-energy photoionization cross section of:
the ground state of  O~{\sc vi} in an 11-state approximation
\citep{cha00}; O~{\sc i} taking into account core relaxation
effects and the Auger smearing of the K edge \citep{gor00b}; and
in its relativistic Breit--Pauli mode, O~{\sc i}--O~{\sc vi} where
wavelengths and $f$-values are listed for the $n=2$ K$\alpha$
resonance complexes with $f>0.1$ \citep{pra03}.

We are interested in a realistic modelling of the oxygen
K edge and in formally establishing its diagnostic potential. For
this purpose we have computed as accurately as possible the
required atomic data, some of which have not been previously
available for the whole oxygen isonuclear sequence, incorporating
them into the {\sc xstar} photoionization code \citep{kal01}.
Previous models have been fitted with two absorption edges and
five Gaussian absorption lines \citep{jue04}. The present data
sets have been computed in a systematic approach previously used
for the iron isonuclear sequence \citep{bau03, pal03a, pal03b,
bau04, men04} which has allowed the determination of the
efficiency of Fe K line emission and absorption in photoionized
gases \citep{kal04}. Extensive comparisons are performed with
several approximations and previous data sets in order to size up
contributing effects and to estimate accuracy ratings. With the
improved opacities, the edge morphology dependency on the
ionization parameter is studied for the first time.


\section{Numerical methods}
The numerical approach has been fully described in \citet{bau03}.
The atomic data, namely level energies, wavelengths, $gf$-values,
radiative widths, total and partial Auger widths, and total and
partial photoionization cross sections are computed with a
portfolio of publicly available atomic physics codes: {\sc
autostruture} \citep{bad86, bad97}, {\sc hfr} \citep{cow81}, and
{\sc bprm} \citep{ber87,sea87}. Although relativistic corrections
are expected to be negligible for the lowly ionized species of the
O sequence, they can be more conspicuous for members with greater
effective charge $z=Z-N+1$, $Z$ being the atomic number and $N$
the electron occupancy. For consistency, wavefunctions are
calculated unless otherwise stated with the relativistic
Breit--Pauli Hamiltonian
\begin{equation}
  \label{hbp}
  H_{\rm bp} = H_{\rm nr} + H_{\rm 1b} + H_{\rm 2b}
\end{equation}
where $H_{\rm nr}$ is the usual non-relativistic Hamiltonian. The
one-body relativistic operators
\begin{equation}
   \label{h1b}
   H_{\rm 1b} = \sum_{n=1}^{N} {f_n({\rm mass}) + f_n({\rm d}) + f_n({\rm so})}
\end{equation}
represent the spin--orbit interaction, $f_n({\rm so})$, the
non-fine-structure mass variation, $f_n({\rm mass})$, and the
one-body Darwin correction, $f_n({\rm d})$. The two-body Breit
operators are given by
\begin{equation}
   \label{h2b}
   H_{\rm 2b} = \sum_{n<m} g_{nm}({\rm so}) + g_{nm}({\rm ss}) + g_{nm}({\rm css})
   + g_{nm}({\rm d}) + g_{nm}({\rm oo})
\end{equation}
where the fine-structure terms are $g_{nm}(so)$ (spin--other-orbit
and mutual spin-orbit), $g_{nm}(ss)$ (spin--spin), and the
non-fine-structure counterparts $g_{nm}(css)$ (spin--spin
contact), $g_{nm}(d)$ (two-body Darwin), and $g_{nm}(oo)$
(orbit--orbit).

{\sc bprm} is based on the close-coupling approximation whereby
the wavefunctions for states of an $N$-electron target and a
colliding electron with total angular momentum and parity $J\pi$
are expanded in terms of the target eigenfunctions
\begin{equation}\label{cc}
  \Psi^{J\pi}={\cal A}\sum_i \chi_i{F_i(r)
  \over r}+\sum_jc_j\Phi_j\ .
\end{equation}
The functions $\chi_i$ are vector coupled products of the target
eigenfunctions and the angular components of the incident-electron
functions; $F_i(r)$ are the radial part of the continuum
wavefunctions that describe the motion of the scattered electron,
and $\mathcal{A}$ is an antisymmetrization operator. The functions
$\Phi_j$ are bound-type functions of the total system constructed
with target orbitals. Breit--Pauli relativistic corrections have
been implemented in {\sc bprm} by \citet{sco80} and \citet{sco82},
but the inclusion of the two-body terms (see Eq.~\ref{h2b}) is at
present in progress. Auger and radiative dampings are taken into
account by means of an optical potential \citep{rob95, gor96,
gor00a} where the resonance energy with respect to the threshold
acquires an imaginary component. In the present work, the
$N$-electron targets are represented with all the fine structure
levels within the $n=2$ complex, and level energies are set to
those obtained in approximation AS2. Positions for states of the
$(N+1)$-electron system are obtained from the peaks in the
photoionization cross sections. This data set is referred to as
RM1.

Core relaxation effects (CRE) are investigated with {\sc
autostruture} by comparing an ion representation where all the
electron configurations have a common basis of orthogonal
orbitals, to be referred to hereafter as approximation AS1, with
one where each configuration has its own set, approximation AS2.
Data sets computed with {\sc hfr} are labelled HF1.


\section{Energy levels and wavelengths}
We have looked into several interactions that can influence the
accuracy of the atomic data, namely configuration interaction
(CI), relativistic corrections, and CRE. It has been found by
calculation that CI with the $n=3$ complex gives rise to small
contributions that can be generally neglected. Computations are
carried out in intermediate coupling that take into account
relativistic corrections even though the latter are small in
particular for low ionization stages. Some effort is then focused
in characterizing CRE.

 Energies calculated with approximations AS1, AS2, HF1, and RM1
for levels within the $n=2$ complex of O ions are listed in
Table~1. Since the accuracy of the transition data, e.g.
wavelengths and $gf$-values, depends on the representations of
both the lower (valence) and upper (K-vacancy) levels, detailed
comparisons are carried out for each type. For the valence levels
(see Table~2), a comparison of AS1 and AS2 reveals the importance
of CRE which lower energies by as much as 20\%. Since RM1 also
neglects CRE, the best accord is with AS1: within 15\% although
the RM1 energies are on average lower by 4\%. It may be noted that
the RM1 calculation for O~{\sc i} has been performed in $LS$
coupling due to the large size of the relativistic option. AS2 and
HF1 agree to 5\% and within 10\% with the spectroscopic values
\citep{moo98} which we find satisfactory.

Regarding K-vacancy levels, the energy differences of the
approximations considered with respect to AS2 are plotted in
Fig~1. The comparison with AS1 shows that, for species with a
half-filled L shell or with greater electron occupancies ($6\leq
N\leq 8$ and $E<544$ eV), CRE again lower energies but by amounts
that increase with $N$: $\Delta E\equiv E(AS1)-E(AS2)< 0.7$ eV for
$N=6$ to $\Delta E\approx 2.5$ eV for $N=8$. On the other hand,
for ions with a half-empty L shell or with lower occupancies
($2\leq N\leq 5$ and $E> 544$ eV), level energies are raised by
CRE particularly for $N=3$ where $\Delta E\gtrsim -1.35$ eV. RM1
behaves, as expected, in a similar fashion to AS1, but some
inconsistencies (e.g. lowered rather than raised levels) occur for
$N<6$ perhaps due to resolution problems in determining the
position of narrow resonances in the photoionization curves. The
agreement between the AS2 and HF1 energies is a sound $\pm 0.6$ eV
which sets a limit to the accuracy level that we can attain, and
encourages us to adopt AS2 and HF1 as our working approximations.

AS2 and HF1 K-vacancy level energies are compared with experiment
and other theoretical data in Table~3. It may be noted that there
are no reported measurements for species with $4\leq N\leq 6$. HF1
level energies are within 0.5 eV of the spectroscopic values for
$N\leq 3$ and the CI values by \citet{cha00} for $N=2$ while AS2
is undesirably higher (0.9 eV) for the $[1\rm{s}]2\rm{p}\
^3\rm{P}^{\rm o}_{1}$ level in O~{\sc vii}. A convincing
comparison with experiment for $N\geq 7$ is impeded by the scatter
displayed in measurements which is well outside the reported error
bars, and the photoionization measurements by \citet{sto97} are
1~eV lower than HF1. The energy positions computed by
\citet{pra03} with {\sc bprm} are noticeably higher than present
results for $N>4$, in particular for the B-like system ($N=5$)
where discrepancies as large as 5 eV are encountered. The values
listed by \citet{mcl98} for O~{\sc ii} ($N=7$), which are also
distinctively high ($\sim$3 eV), correspond to the CI target they
prepared for the non-relativistic $R$-matrix calculation of the
photoionization of O~{\sc i}. They contrast with those quoted for
an equivalent target by \citet{gor00b} which are in good agreement
with \citet{sto97}. However, the value obtained by \citet{mcl98}
for the $[1\rm{s}]2\rm{p}^5\ ^3\rm{P}^{\rm o}$ hole state in
O~{\sc i} is in good agreement with AS2 and HF1.

CRE also have an impact on the transition wavelengths where, as
depicted in Fig.~2, the comparison of AS1 and RM1 with AS2
indicates that in general CRE lead to increasingly longer
wavelength for $N\geq 6$ ($\lambda > 23$ \AA) and shorter for
higher charge states. It may also be seen in Fig.~2 that
wavelengths obtained with AS2 and HF1 agree to $\pm 0.02$ \AA\
setting again an accuracy bound. (AS2 and HF1 wavelengths are also
listed in Table~4.) The present wavelengths are compared with
experiment and other theoretical predictions in Table~5. For
$N\leq 4$ the present data, in particular HF1, agree with
measurements and \citet{vai78} to $\pm 0.02$ \AA; in this respect,
for $2<N\leq 4$ the values by \citet{che85, che86} are somewhat
long while those by \citet{beh02} are short, and for $N=3$ the
values by \citet{pra03} are also long. For $5\leq N\leq 6$ the
outcome is less certain due to absence of laboratory measurements
and inconsistencies in the computed values, e.g. the order of the
upper levels in $N=5$. For $N>6$ comparisons with measurements
seem to indicate that all theoretical results are slightly short
although the agreement of HF1 is still within $\sim$0.02 \AA\
except for the transition from the ground level of O~{\sc ii} and
the measurement by \citet{sto97} in O~{\sc i} which are just out
(less than 0.05\AA). It must be pointed out again the undesirable
scatter of the experimental data for O~{\sc i}.


\section{$gf$-values}
In the following discussion of $gf$-values, the transitions
involving the strongly mixed $[1\rm{s}]2\rm{p}^4\ ^2\rm{D}_{3/2}$
and $^2\rm{P}_{3/2}$ levels in O~{\sc ii} are excluded due to
extensive cancellation that result in completely unreliable data;
also the $1\rm{s}^2\ ^1\rm{S}_{0} \rightarrow [1\rm{s}]2\rm{p}\
^3\rm{P}^{\rm o}_{1}$ intercombination transition in O~{\sc vii}
with a very small $gf$-value ($gf<10^{-3}$) that would require a
much greater effort to guarantee respectful accuracy.

A comparison of $gf$-values obtained with the AS1 and AS2
approximations shows that CRE are relatively unimportant, leading
to differences much less than 10\% except for transitions
involving the four $[1\rm{s}]2\rm{s}2\rm{p}\ ^2\rm{P}^{\rm o}$
relatively mixed levels of O~{\sc vi} where they jump up to
$\sim$15\%. In Fig.~3 we compare the AS2 and HF1 $gf$-values (also
listed in Table~4) showing an agreement within 10\% except for
O~{\sc i} where the latter are consistently 25\% higher. In a
comparison of AS2 and HF1 $f$-values with other theoretical data
sets in Table~6, it is found that those by \citet{pra03} are
generally 20\% smaller; however, an interesting point arises
regarding the discrepancies between AS2 and HF1 where
\citet{pra03} favors AS2 for O~{\sc i} and HF1 for the small
$f$-value in O~{\sc vi} (see Table~6).


\section{Radiative and Auger widths}
The radiative width of the $j$th level is defined as
\begin{equation}
A_j=\sum_i A_{ji}
\end{equation}
where $A_{ji}$ is the rate (partial width) for a downward ($j>i$)
transition. It is found by calculation that CRE disturb the
radiative widths of K-vacancy levels in the O isonuclear sequence
by 10\% or less (5\% for $N>5$) except when strong admixture
occurs, namely for $[1\rm{s}]2\rm{p}^4\ ^2\rm{D}_{3/2}$ and
$^2\rm{P}_{3/2}$ in O~{\sc ii} and the four
$[1\rm{s}]2\rm{s}2\rm{p}\ ^2\rm{P}^{\rm o}$ levels in O~{\sc vi}
where differences are around 20\%.

By contrast, as shown in Fig.~4, CRE reduce Auger rates by amounts
that grow with $N$: from 10\% for $N=4$ where $\log A_{\rm a}< 14$
up to 35\% for $N=8$; however, this trend is broken for $N=3$ due
to strong level mixing where those for the
$[1\rm{s}]2\rm{s}2\rm{p}(^3{\rm P}^{\rm o})\ ^2\rm{P}^{\rm o}_J$
are increased by a factor of 2 (not shown in Fig.~4) even though
the $[1\rm{s}]2\rm{s}2\rm{p}(^1{\rm P}^{\rm o})\ ^2\rm{P}^{\rm
o}_J$ levels are hardly modified (less than 5\%). The Auger widths
obtained with HF1 for the former levels are $\sim$20\% higher that
AS2 while the remaining, as shown in Fig.~4, are on average lower
by 10\%. With this outcome we estimate an accuracy rating for the
Auger rates at just better than 20\%. As further support for this
assertion, in Table~7 we compare our branching ratios for KLL
Auger transitions with other theoretical and experimental values.
It may be seen that the AS1, AS2, and HF1 data are stable to
better than 10\%, and the accord with other data sets is for most
cases better than 20\%. Radiative and Auger widths computed in
approximations AS2 and HF1 are listed in Table~8.


\section{Photoionization cross sections}

High-energy photoionization cross sections of O parent ions with
electron occupancies $3\leq N\leq 8$ have been computed in
approximation RM1. Intermediate coupling has been used throughout
except for O~{\sc i} because of computational size and negligible
relativistic corrections. Following \citet{gor99} and
\citet{gor00b}, damping is taken into account in detail in order
to bring forth the correct K threshold behavior. In Fig.~5 our
cross sections are compared with those computed by \citet{pra03}
with the {\sc bprm} package in the region near the $n=2$
resonances and those by \citet{rei79} in a central field potential
where it may be seen that, even though the background cross
sections are in satisfactory accord, the positions and sharpness
of the edges are discrepant. In the present approach, due to the
astrophysical interest of both the $n=2$ resonances and the edge
region, the resonance structure is treated in a unified manner
showing that a sharp K edge occurs only in O~{\sc vi} due to the
absence of spectator Auger (KLL) channels of the type
\begin{eqnarray}
[\rm{1s}]\rm{2p}^\mu n\rm{p}& \rightarrow & \rm{2p}^{\mu-2}n\rm{p}+\rm{e}^-\\
                            & \rightarrow & [\rm{2s}]\rm{2p}^{\mu-1}n\rm{p}+\rm{e}^-\\
                            & \rightarrow & [\rm{2s}]^2\rm{2p}^{\mu}n\rm{p}+\rm{e}^-\ .
\end{eqnarray}
For the remaining ions, these KLL processes dominate over
participator Auger (KL$n$) decay
\begin{eqnarray}
[\rm{1s}]\rm{2p}^\mu n\rm{p}& \rightarrow & \rm{2p}^{\mu-1}+\rm{e}^-\\
                            & \rightarrow & [\rm{2s}]\rm{2p}^{\mu}+\rm{e}^-
\end{eqnarray}
causing edge smearing by resonances with symmetric profiles of
nearly constant width. Participator Auger decay, on the other
hand, gives rise to the familiar Feschbach resonances---displayed
by O~{\sc vi} in the region below the K threshold (see
Fig.~5)---whose widths become narrower with $n$ thus maintaining
sharpness over the spectral head.

In order to elucidate further the interesting properties of K
resonances, we analyze the experimental and theoretical
predictions reported for the two
$[1\rm{s}]2\rm{p}^4(^{4}\rm{P})n\rm{p}$  and
$[1\rm{s}]2\rm{p}^4(^{2}\rm{P})n\rm{p}$ Rydberg series in O~{\sc
i} that exhibit significant discrepancies. Regarding resonance
positions, a stringent test is to compare quantum defects, $\mu$,
defined as
\begin{eqnarray}
E= E_{\rm lim} - \frac{z^2}{(n-\mu)^2}
\end{eqnarray}
where $E_{\rm lim}$ is the series-limit energy in Rydbergs;
$z\equiv Z-N+1$ is the effective charge ($z=1$ for O~{\sc i}); and
$n$ is the principal quantum number. In Fig.~6, the RM1 quantum
defects for these two series are plotted relative to their
respective series limits and compared with those obtained in
photoionization measurements \citep{sto97} and the $R$-matrix
calculation of \citet{mcl98}. Firstly, it may be seen that the
experimental error bars grow sharply with $n$; hence greater
accuracy for high $n$ is obtained by quantum-defect extrapolation
than actual measurement. Secondly, the agreement of experiment
with RM1 is very good while that with \citet{mcl98}, who used
practically the same numerical method ($R$-matrix), is only a poor
40\%. This outcome may be due to their neglect of damping effects
or their choice of a Hartree--Fock basis for the 1s, 2s, and 2p
orbitals which results in a deficient representation for the
K-vacancy states. Since the $R$-matrix method employs a common set
of target orbitals for both valence and K-vacancy states, we have
found the most adequate orbital basis to be that optimized
variationally with an energy sum comprising all the states in the
target model. The importance of K-vacancy state representations
has also been stressed by \citet{gor00b}. Thirdly, it is found
that the RM1 quantum defects ($n\geq 3$) are hardly modified by
small target-energy shifts prior to Hamiltonian diagonalization, a
standard $R$-matrix procedure to enhance accuracy by using
experimental thresholds; therefore, the absolute energy positions
of these resonances mostly depend on the actual position of their
series limits. Moreover, as shown in Fig.~7, the huge and
incorrect quantum-defect decline obtained for high $n$ in
\citet{mcl98} when using two different sets of experimental
thresholds, namely those of \citet{kra94} and \citet{sto97}, are
not due in our opinion to the threshold-energy shifts but to
resonance misidentifications where $\mu(n+1)-\mu(n)\approx -1$
instead of approximately zero (see Fig.~7). The quantum-defect
changes for $n\leq 6$ in \citet{mcl98} as thresholds are shifted
are very small in accord with our findings. By contrast,
experimental quantum defects are markedly sensitive to the
series-limit positions; for instance, if the quantum defects in
the photoionization experiment of \citet{men96} are computed with
the series limits by \citet{kra94}, they display noticeable
discrepancies (see Fig.~7) while if they are computed with those
by \citet{sto97} (not shown), consensus is reached.

It is inferred from the above comparison that the energy positions
obtained by Auger electron spectrometry \citep{kra94, cal94} for
the $[1\rm{s}]2\rm{p}^4\ ^4\rm{P}$ and $^2\rm{P}$ states of O~{\sc
ii} and listed in Table~3 should perhaps be scaled down by 0.4 eV.
If this proposition is extended to the $[1\rm{s}]2\rm{p}^5\
^3\rm{P}^{\rm o}$ state of O~{\sc i}, then the measurements by
\citet{sto97} and \citet{kra94} are brought to close agreement;
the only remaining experimental discrepancy for the latter state
would then be the measurement by \citet{men96} still high by 1 eV.
Furthermore, contrary to the
$[1\rm{s}]2\rm{p}^4(^{4}\rm{P})n\rm{p}\ ^3\rm{P}^{\rm o}$
resonances of O~{\sc i} in our RM1 calculation, the absolute
energy position of $[1\rm{s}]2\rm{p}^5\ ^3\rm{P}^{\rm o}$ is
insensitive to threshold-energy shifts. This is a consequence of
its wavefunction being mostly represented by the second term of
the close-coupling expansion (\ref{cc}). Therefore, a useful
parameter in comparisons with experiment---in as much as it gives
an indication of both a balanced close-coupling expansion and
experimental consistency---is the energy interval between the
lowest $n=2$ and $n=3$ resonances, $\Delta E(2,3)$ say. In
Table~9, the RM1 $\Delta E(2,3)$ for the whole O isonuclear
sequence are tabulated and compared with experimental and
theoretical estimates. The 0.07 Ryd (1 eV) discrepancy in the
$\Delta E(2,3)$ obtained from the measurements by \citet{sto97}
and \citet{men96} in O~{\sc i} ($N=8$) denotes irregular data:
since the positions for the $(^{4}\rm{P})3\rm{p}$ state differ by
only 0.07 eV the issue is then the $[1\rm{s}]2\rm{p}^5\
^3\rm{P}^{\rm o}$. The agreement of RM1 with \citet{sto97} is very
good; however, if the RM1 $(^{4}\rm{P})$ threshold is improved by
replacement with the experimental value at Halmiltonian
diagonalization, then $\Delta E(2,3)$ is reduced to 1.00. In other
words, the introduction of experimental thresholds improves the
positions of resonances with $n\geq 3$ relative to the ionization
threshold but not those within $n=2$, as previously mentioned. It
may be appreciated in Table~9 that this finding is corroborated by
the $\Delta E(2,3)$ resulting from the threshold-energy shifting
in \citet{mcl98}. It is also seen in Table~9 that for the O
isonuclear sequence $\Delta E(2,3)\approx z$ for $N\geq 2$ and
$\Delta E(2,3)\approx z+1$ for $N=1$ if $\Delta E(2,3)$ is
expressed in Ryd.

Regarding K-resonance widths in O~{\sc i}, from the Auger rates
listed in Table~8 widths of 159 meV (AS2) and 163 meV (HF1) are
obtained for $[1\rm{s}]2\rm{p}^5\ ^3\rm{P}^{\rm o}$. These are in
good accord ($\sim$15\%) with the measurements of 140 meV
\citep{kra94, men96} and 142 eV \citep{sto97} and the theoretical
estimates of 169 meV \citep{sah94} and 139 meV \citep{pet94}. The
value of 185.22 meV listed by \citet{mcl98}, on the other hand, is
somewhat high. These authors also predict for the
$[1\rm{s}]2\rm{p}^4(^{4}\rm{P})n\rm{p}\ ^3\rm{P}^{\rm o}$ widths a
decrease with $n$. As previously mentioned, the utter dominance of
KLL decay (see Eqs~7--9) gives rise to constant resonance widths
confirmed with AS1 for $3\leq n\leq 5$ to be at 103 meV and with
KL$n$ branching ratios of 5 parts in 1000 or less.


\section{Opacities}

In order to investigate the variations of the oxygen K-edge
structure with plasma conditions, our best atomic data for O ions
have been incorporated in the {\sc xstar} modelling code
\citep{kal01}. Opacities for ionization parameters in the range
$0.001 \leq \xi \leq 10$ have then been obtained with a simple
model denoted by the following specifications: gas density of
$10^{12}$ cm$^{-3}$; X-ray luminosity source of $10^{44}$ erg/s;
and solar abundances for H, He, and O. Since the density is
assumed constant, the well-known expression for the ionization
parameter
\begin{equation}
   \label{ip}
     \xi = L / n R^2
\end{equation}
is implemented where $L$ is the luminosity of the source, $R$ its
distance, and $n$ the density of the gas \citep{tar69}. Therefore,
the monochromatic opacities are calculated assuming that
ionization and heating are mainly due to an external source of
continuum photons.

In Fig.~\ref{opa} opacities for different values of $\log\xi$ are
plotted in the 400--800 eV energy range showing a rich variability
mainly determined by the resonance structure and K edges of the
abundant ionic species. For a lowly ionized plasma ($\log\xi=-3$),
the opacity is dominated by the O~{\sc i} edge at $\sim$545 eV and
the K$\alpha$ resonance structure belonging to O~{\sc i} and
O~{\sc ii} in the 530--540 eV energy interval. For $\log\xi=-2$
the O~{\sc ii} edge is also conspicuous at $\sim$565 eV, but by
$\log\xi=-1.5$ is almost vanished being replaced by the O~{\sc iii}
and O~{\sc iv} edges at $\sim$600 eV and $\sim$630 eV, respectively.
While the latter is still distinctive for $\log\xi=-1$, the O~{\sc
v} edge arises at $\sim$665 eV and becomes the dominant feature
for $\log\xi=0$. At a high degree of ionization ($\log\xi=1$), a
three-edge structure (O~{\sc v}, O~{\sc vi}, and O~{\sc vii}) is
observed together with strong absorption features due to K$\beta$ and
high-$n$ resonances that in general enhance edge smearing, the opposite
characteristics of low ionization. This behavior at the two extremes
of the ionization parameter is similar to that reported for the
Fe isonuclear sequence by \cite{pal02}, an important difference
being that the oxygen opacities show important absorption features
in the edge region due to the always open L shell.


\section{Discussion and conclusions}
Extensive computations of the atomic data required for the
spectral modelling of K-shell photoabsorption of oxygen ions have
been carried out in a multi-code approach. This method has proven
to be---as in a previous systematic treatment of the Fe isonuclear
sequence \citep{bau03, pal03a, pal03b, bau04, men04}---effective
in sizing up the relevant interactions and in obtaining accuracy
rating estimates, the latter often overlooked in theoretical work.
Inter-complex CI and relativistic corrections are found to be
small, but CRE are shown to be conspicuous in most processes
involving K-vacancy states. The inclusion of CRE in ion
representations is therefore compelling which in practice is
managed (e.g. in the {\sc hfr} and {\sc autostructure} codes) by
considering non-orthogonal orbital bases for K-vacancy and valence
states; otherwise (e.g. {\sc bprm}) orthogonal orbital bases are
optimized by minimizing energy sums that span both types of
states.

By considering several approximations and comparing with reported
experimental and theoretical data, the present level energies and
wavelength for O ions with electron occupancies $N\leq 4$ are
estimated to be accurate to within 0.5 eV and 0.02 \AA,
respectively. For ions with $N>4$, the absence of measurements in
particular for $5\leq N\leq 6$, the wide experimental scatter for
$N>6$, and discrepancies among theoretical data sets make the
estimate of reliable accuracy ratings less certain. It is found
that for $N>4$ our level energies (wavelengths) are probably
somewhat high (short) by up to 1.5 eV (0.05 \AA) in neutral
oxygen. Our best results are from approximation HF1.

The radiative data ($gf$-values and total widths) computed with
approximations AS2 and HF1 are expected to be accurate to 10\%
except in O~{\sc i} where the rating is not better than 25\% and
in transitions involving strongly mixed levels, namely the
$[1\rm{s}]2\rm{p}^4\ ^2\rm{D}_{3/2}$ and $^2\rm{P}_{3/2}$ levels
in O~{\sc ii} and to a lesser extent the four
$[1\rm{s}]2\rm{s}2\rm{p}\ ^2\rm{P}_J^{\rm o}$ levels in O~{\sc
vi}, where it is unreliable. Admixture in the latter levels also
affect Auger rates which otherwise are expected to be accurate to
20\%. The complete and ranked set of radiative and Auger data for
transitions involving K-vacancy levels in O ions is one of the
main achievements of the present work.

A second contribution is the high-energy photoabsorption cross
sections for ions with $N>3$ that depict in detail the resonance
structure of the K edge. Such structure has been shown to be
dominated by KLL damping (often neglected in previous work) which
gives the edge a blunt imprint with arguably diagnostic potential.
A comparison with the experimental K resonance positions in O~{\sc
i} results in conclusive statements about the importance of
$R$-matrix computations with well balanced close-coupling
expansions, i.e. where both terms of expression (\ref{cc}) are
separately convergent. In this respect, the accuracy of the energy
interval between the $n=2$ and $n=3$ components of a K resonance
series gives a measure of the close-coupling balance, and for the
O isonuclear sequence, this energy interval when expressed in Ryd
has been shown to be close to the effective charge, $z\equiv
Z-N+1$, for members with $N\geq 2$ and near $z+1$ for $N=1$. This
finding could be exploited in spectral identification.

The present atomic data sets have been incorporated in the {\sc
xstar} modelling code in order to generate improved opacities in
the oxygen K-edge region using simple constant density slab
models. Their behavior as a function of ionization parameter,
shown in detail for the first time, is similar to that of the iron
K edge with the exception of omnipresent K$\alpha$ resonances.
Finally, as a continuation to the present work we intend to treat
with the same systematic methodology the Ne isonuclear sequence
and then other sequences of astrophysical interest (Mg, Si, S, Ar,
and Ca), always bearing in mind the scanty availability of
laboratory measurements which as shown in the present work are
unreplaceable requisites in theoretical fine-tuning.


\acknowledgments MAB acknowledges partial support from FONACIT,
Venezuela, under contract No. S1-20011000912. TWG was supported in
part by the NASA Astronomy and Physics Research and Analysis
Program. PP was supported by a return grant of the Belgian Federal
Scientific Policy.



\clearpage
\begin{figure}
\epsscale{.80}
\plotone{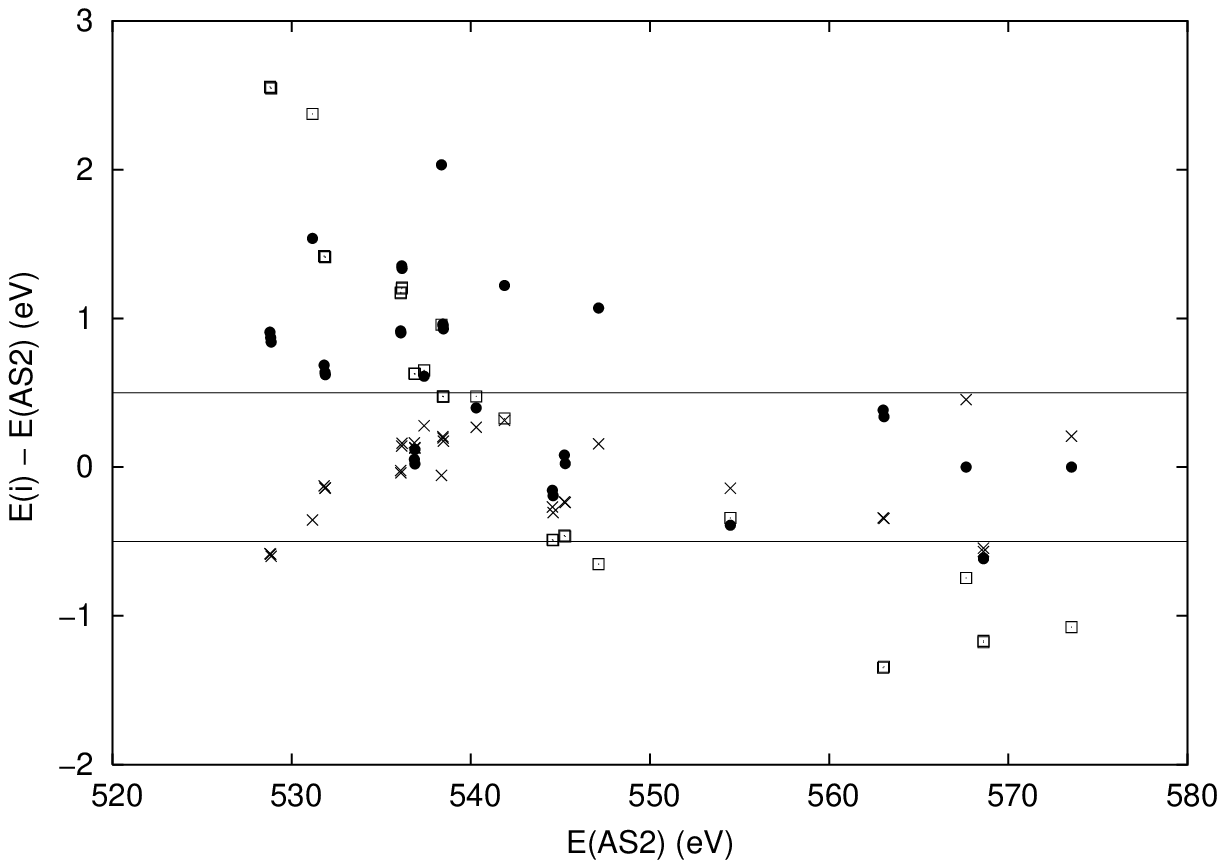}
\caption{Energy differences for K-vacancy levels in the O isonuclear
sequence between those computed with approximation AS2 and those with
AS1 (open squares), HF1 (crosses) and RM1(filled circles). Differences
with AS1 and RM1 are due mainly to CRE. The agreement between AS2 and
HF1 is within $\pm 0.5$ eV.}
\end{figure}


\begin{figure}
\epsscale{.80}
\plotone{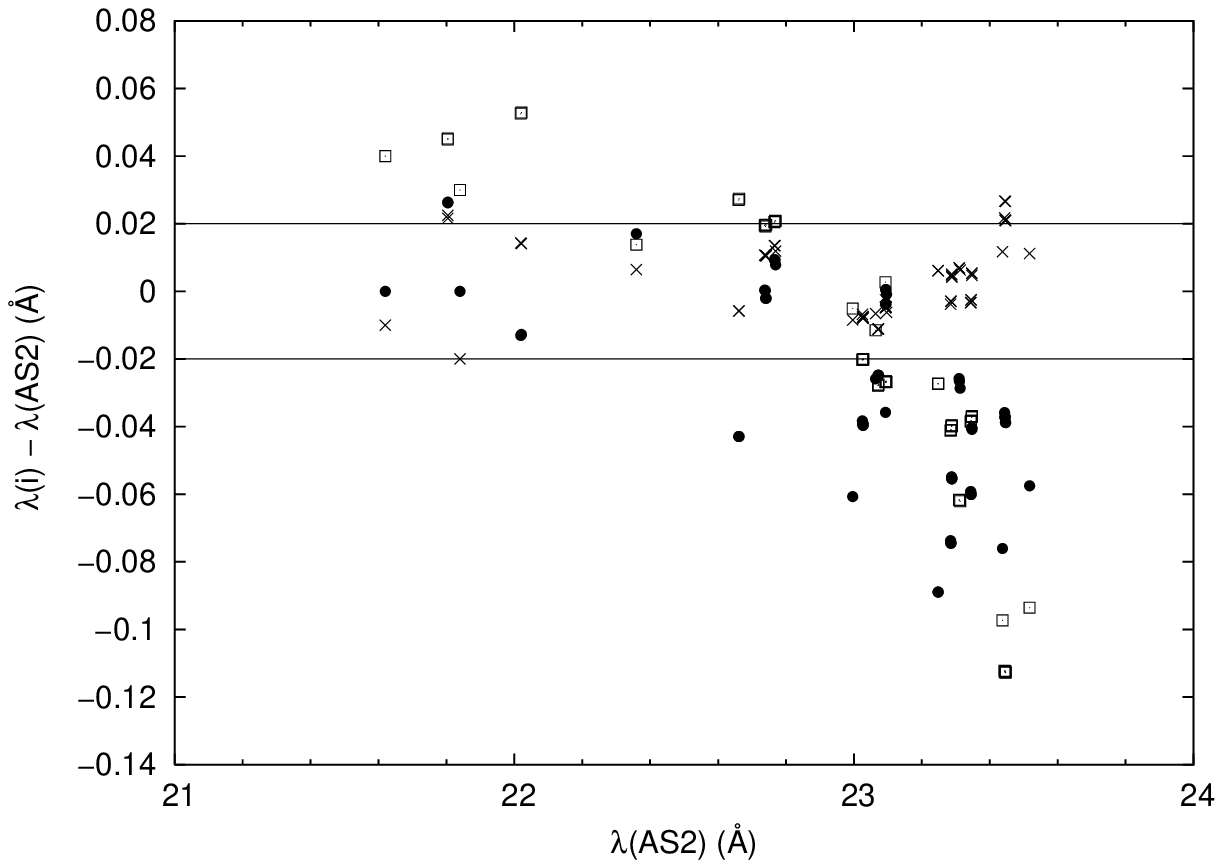}
\caption{Wavelength differences for K$\alpha$ transitions in the O
isoelectronic sequence between those computed with approximation AS2 and those
with AS1 (open squares), HF1 (crosses), and RM1 (solid circles). The agreement
between AS2 and HF1 is within $\pm 0.02$ \AA.}
\end{figure}

\begin{figure}
\epsscale{.80}
\plotone{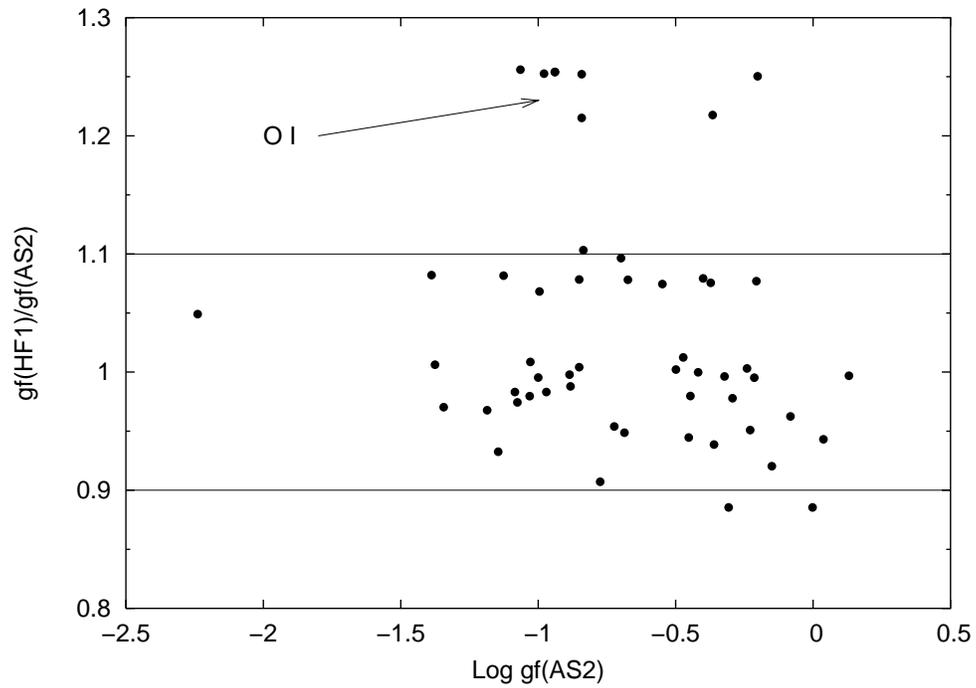}
\caption{Comparison of $gf$-values for K$\alpha$
transitions in O ions computed with approximations AS2 and HF1. It
may be seen that, while the agreement for most transitions is
within 10\%, the HF1 $gf$-values for O~{\sc i} are consistently
larger by 25\%.}
\end{figure}


\begin{figure}
\epsscale{.80}
\plotone{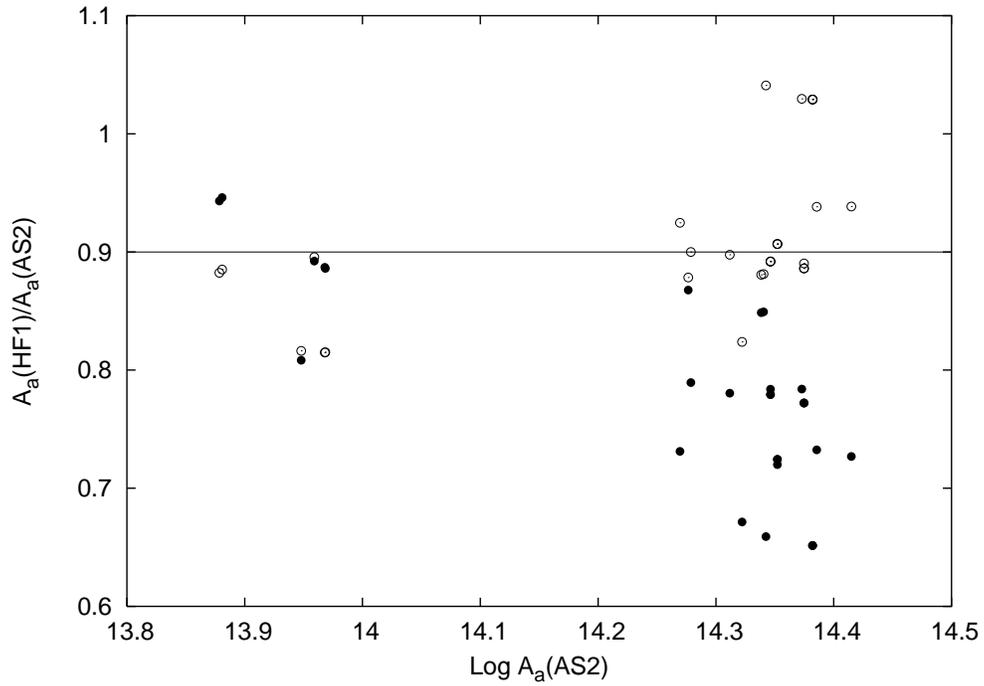}
\caption{Comparison of Auger widths (s$^{-1}$)
for K-vacancy levels in O ions computed with approximation AS2
and those with AS1 (filled circles) and HF1 (open circles).
Differences between AS1 and AS2 (as large as 35\%) are due to CRE.
It may also be seen that HF1 are on average 10\% lower than AS2.}
\end{figure}


\begin{figure}
\epsscale{.80}
\plotone{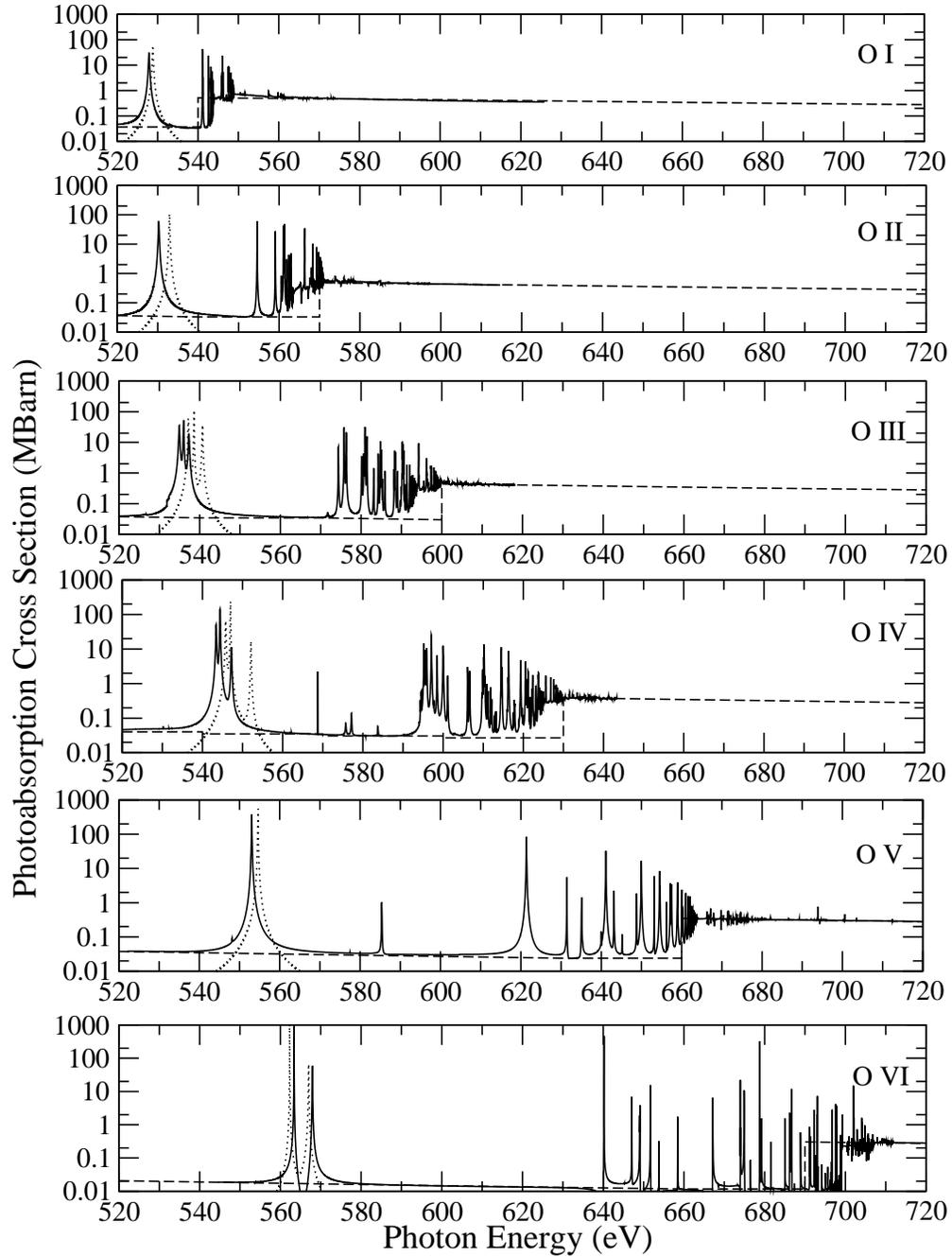}
\caption{High-energy photoionization cross sections of O ions showing
the structure of the K edge. Solid curve --- RM1; dotted curve --- \citet{pra03};
broken curve --- \cite{rei79}.}
\end{figure}

\begin{figure}
\epsscale{.80}
\plotone{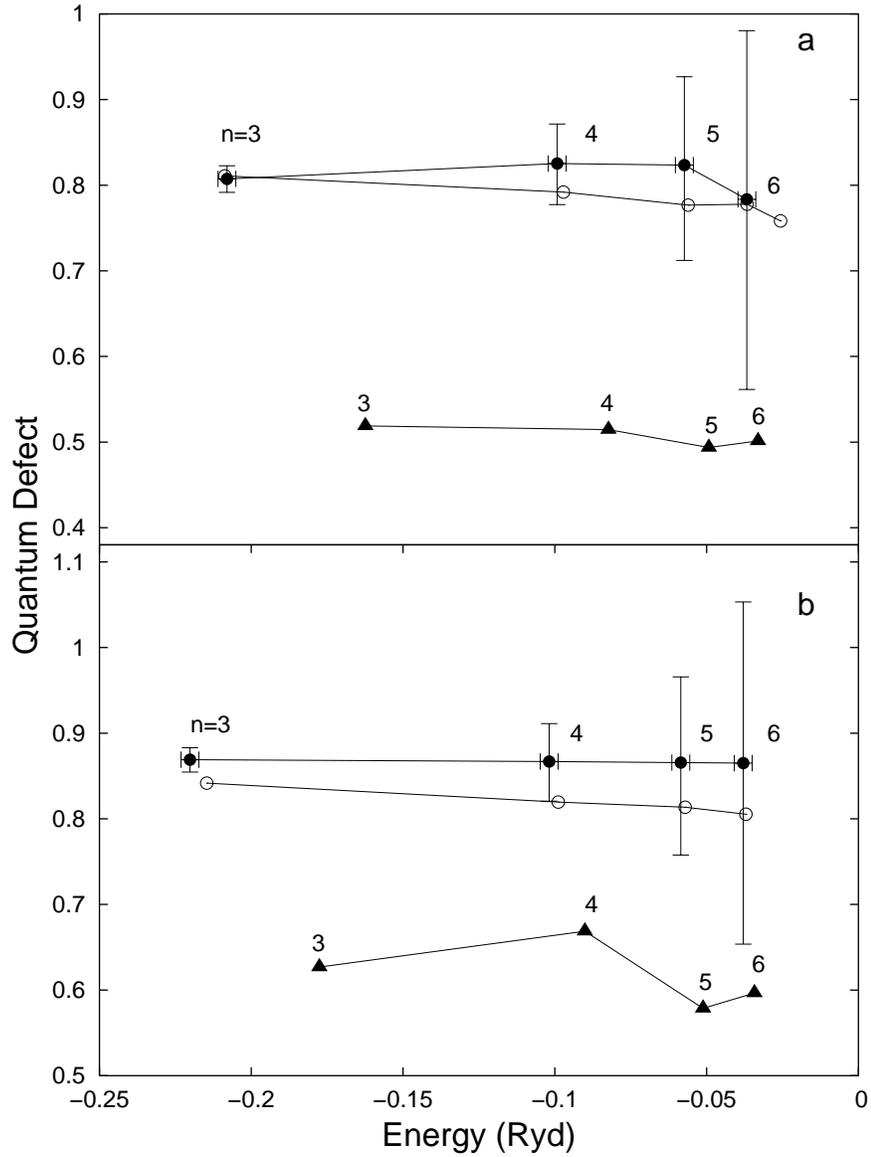}
\caption{Quantum defects for the (a)
$[1\rm{s}]2\rm{p}^4(^4\rm{P})n\rm{p}\ ^3\rm{P}^{\rm o}$ and (b)
$[1\rm{s}]2\rm{p}^4(^2\rm{P})n\rm{p}\ ^3\rm{P}^{\rm o}$ ($3\leq
n\leq 6$) resonance series of O~{\sc i} plotted relative to their
respective threshold energies. Circles --- present RM1 results.
Filled circles --- experiment of \citet{sto97}. Filled triangles
--- $R$-matrix data by \citet{mcl98}.}
\end{figure}

\begin{figure}
\epsscale{.80}
\plotone{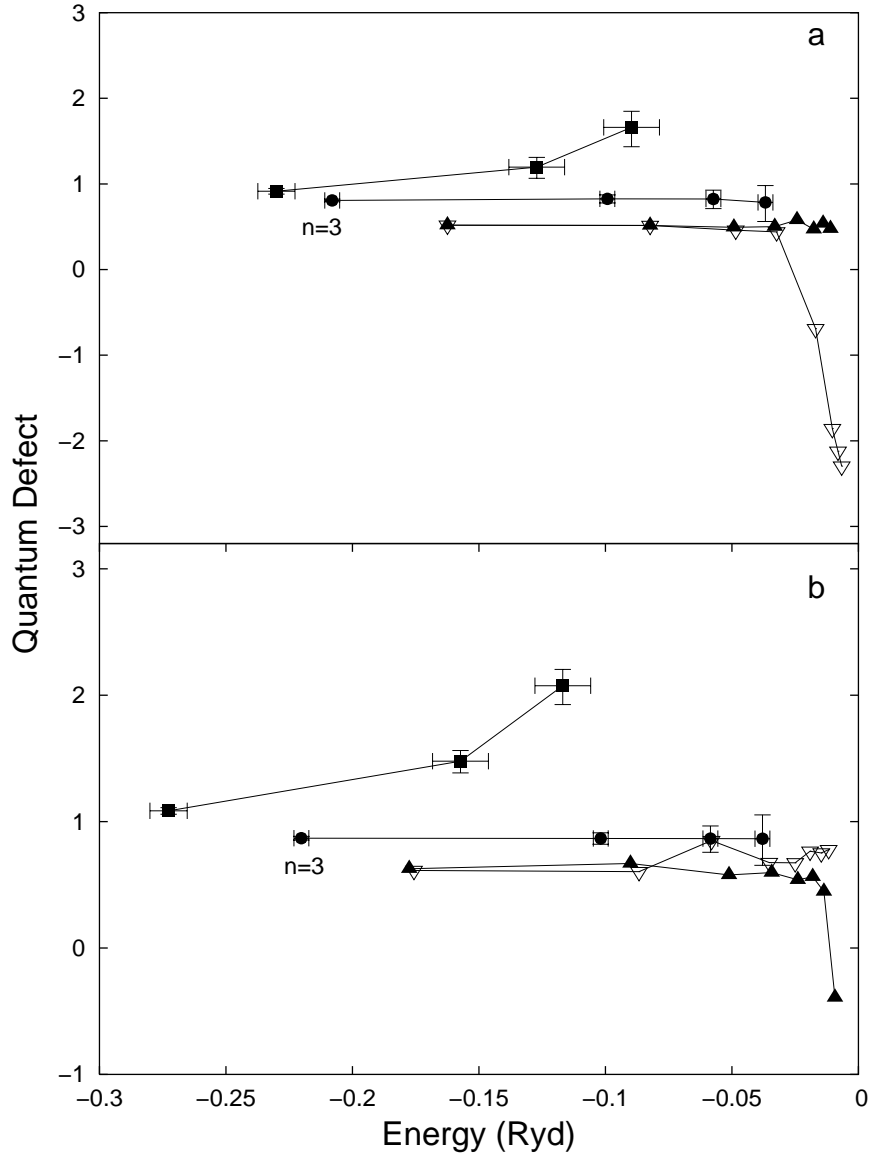}
\caption{Quantum defects for the (a)
$[1\rm{s}]2\rm{p}^4(^4\rm{P})n\rm{p}\ ^3\rm{P}^{\rm o}$ and (b)
$[1\rm{s}]2\rm{p}^4(^2\rm{P})n\rm{p}\ ^3\rm{P}^{\rm o}$ ($3\leq
n\leq 10$) resonance series of O~{\sc i} plotted relative to their
respective threshold energies. Filled squares --- experiment of
\citet{kra94} and \citet{men96}. Filled circles --- experiment of
\citet{sto97}. Triangles
--- $R$-matrix quantum defects of \citet{mcl98} relative to the
experimental thresholds of \citet{kra94}. Filled triangles ---
$R$-matrix quantum defects of \citet{mcl98} relative to the
experimental threshold by \citet{sto97}.}
\end{figure}


\begin{figure}
\label{opa}
\epsscale{.80}
\plotone{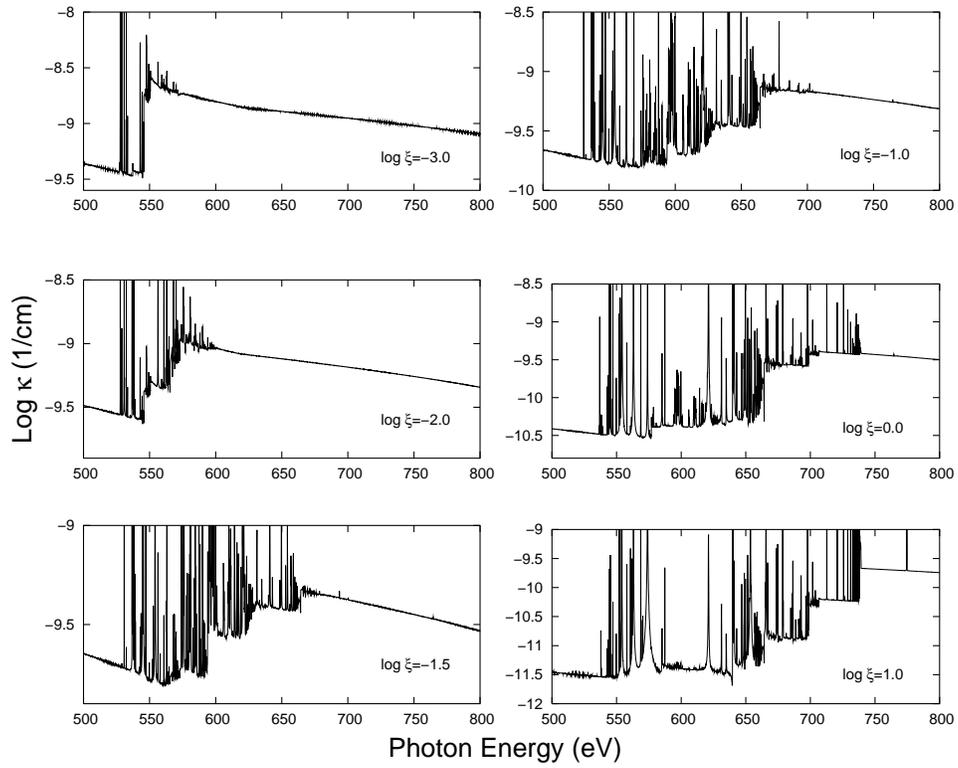}
\caption{Opacities for a photoionized gas in the oxygen K-edge region for
ionization parameters in the range $-3 \leq \log\xi\leq 1$.}
\end{figure}

\clearpage
\begin{deluxetable}{rrlrrrrrrlrrrr}
\tablecolumns{14} \tabletypesize{\scriptsize} \tablewidth{0pc}
\tablecaption{Calculated energies (eV) for valence and K-vacancy
levels in O ions} \tablehead{\colhead{$N$} & \colhead{$i$} &
\colhead{Level} & \colhead{AS1} & \colhead{AS2} & \colhead{HF1}&
\colhead{RM1} &
           \colhead{$N$} & \colhead{$i$} & \colhead{Level} & \colhead{AS1} & \colhead{AS2} & \colhead{HF1}& \colhead{RM1} \\
} \startdata
2 & 1 & $1\rm{s}^2\ ^1\rm{S}_{0}$                         & 0.000 & 0.000 & 0.000 &       & 7 & 1 & $2\rm{p}^3\ ^4\rm{S}_{3/2}^{\rm o}$               & 0.000 & 0.000 & 0.000 & 0.000 \\
2 & 2 & $[1\rm{s}]2\rm{p}\ ^3\rm{P}^{\rm o}_{1}$          & 566.9 & 567.7 & 568.1 &       & 7 & 2 & $2\rm{p}^3\ ^2\rm{D}_{5/2}^{\rm o}$               & 3.972 & 3.711 & 3.770 & 3.615 \\
2 & 3 & $[1\rm{s}]2\rm{p}\ ^1\rm{P}^{\rm o}_{1}$          & 572.5 & 573.5 & 573.7 &       & 7 & 3 & $2\rm{p}^3\ ^2\rm{D}_{3/2}^{\rm o}$               & 3.975 & 3.714 & 3.770 & 3.614 \\
3 & 1 & $2\rm{s}\ ^2\rm{S}_{1/2}$                         & 0.000 & 0.000 & 0.000 & 0.000 & 7 & 4 & $2\rm{p}^3\ ^2\rm{P}_{3/2}^{\rm o}$               & 5.393 & 5.064 & 5.131 & 5.655 \\
3 & 2 & $[1\rm{s}]2\rm{s}2\rm{p}\ ^2\rm{P}_{1/2}^{\rm o}$ & 561.7 & 563.0 & 562.7 & 563.4 & 7 & 5 & $2\rm{p}^3\ ^2\rm{P}_{1/2}^{\rm o}$               & 5.395 & 5.066 & 5.131 & 5.654 \\
3 & 3 & $[1\rm{s}]2\rm{s}2\rm{p}\ ^2\rm{P}_{3/2}^{\rm o}$ & 561.7 & 563.1 & 562.7 & 563.4 & 7 & 6 & $[1\rm{s}]2\rm{p}^4\ ^4\rm{P}_{5/2}$              & 533.2 & 531.8 & 531.7 & 532.5 \\
3 & 4 & $[1\rm{s}]2\rm{s}2\rm{p}\ ^2\rm{P}_{1/2}^{\rm o}$ & 567.4 & 568.6 & 568.0 & 568.0 & 7 & 7 & $[1\rm{s}]2\rm{p}^4\ ^4\rm{P}_{3/2}$              & 533.3 & 531.9 & 531.7 & 532.5 \\
3 & 5 & $[1\rm{s}]2\rm{s}2\rm{p}\ ^2\rm{P}_{3/2}^{\rm o}$ & 567.4 & 568.6 & 568.1 & 568.0 & 7 & 8 & $[1\rm{s}]2\rm{p}^4\ ^4\rm{P}_{1/2}$              & 533.3 & 531.9 & 531.7 & 532.5 \\
4 & 1 & $2\rm{s}^2\ ^1\rm{S}_{0}$                         & 0.000 & 0.000 & 0.000 & 0.000 & 7 & 9 & $[1\rm{s}]2\rm{p}^4\ ^2\rm{D}_{5/2}$              & 537.3 & 536.1 & 536.1 & 537.0 \\
4 & 2 & $[1\rm{s}]2\rm{p}\ ^1\rm{P}_{1}^{\rm o}$          & 554.2 & 554.5 & 554.4 & 554.1 & 7 &10 & $[1\rm{s}]2\rm{p}^4\ ^2\rm{D}_{3/2}$              & 537.3 & 536.1 & 536.1 & 537.0 \\
5 & 1 & $2\rm{p}\ ^2\rm{P}_{1/2}^{\rm o}$                 & 0.000 & 0.000 & 0.000 & 0.000 & 7 &11 & $[1\rm{s}]2\rm{p}^4\ ^2\rm{P}_{3/2}$              & 537.4 & 536.2 & 536.3 & 537.5 \\
5 & 2 & $2\rm{p}\ ^2\rm{P}_{3/2}^{\rm o}$                 & 0.057 & 0.049 & 0.047 & 0.047 & 7 &12 & $[1\rm{s}]2\rm{p}^4\ ^2\rm{P}_{1/2}$              & 537.4 & 536.2 & 536.3 & 537.5 \\
5 & 3 & $[1\rm{s}]2\rm{p}^2\ ^2\rm{D}_{5/2}$              & 544.1 & 544.6 & 544.3 & 544.4 & 7 &13 & $[1\rm{s}]2\rm{p}^4\ ^2\rm{S}_{1/2}$              & 539.3 & 538.4 & 538.3 & 540.4 \\
5 & 4 & $[1\rm{s}]2\rm{p}^2\ ^2\rm{D}_{3/2}$              & 544.1 & 544.6 & 544.3 & 544.4 & 8 & 1 & $2\rm{p}^4\ ^3\rm{P}_{2}$                         & 0.000 & 0.000 & 0.000 & 0.000 \\
5 & 5 & $[1\rm{s}]2\rm{p}^2\ ^2\rm{P}_{1/2}$              & 544.8 & 545.2 & 545.0 & 545.3 & 8 & 2 & $2\rm{p}^4\ ^3\rm{P}_{1}$                         & 0.023 & 0.020 & 0.018 & 0.000 \\
5 & 6 & $[1\rm{s}]2\rm{p}^2\ ^2\rm{P}_{3/2}$              & 544.8 & 545.3 & 545.0 & 545.3 & 8 & 3 & $2\rm{p}^4\ ^3\rm{P}_{0}$                         & 0.033 & 0.029 & 0.027 & 0.000 \\
5 & 7 & $[1\rm{s}]2\rm{p}^2\ ^2\rm{S}_{1/2}$              & 546.5 & 547.1 & 547.3 & 548.2 & 8 & 4 & $2\rm{p}^4\ ^1\rm{D}_{2}$                         & 2.353 & 2.181 & 2.200 & 2.149 \\
6 & 1 & $2\rm{p}^2\ ^3\rm{P}_{0}$                         & 0.000 & 0.000 & 0.000 & 0.000 & 8 & 5 & $2\rm{p}^4\ ^1\rm{S}_{0}$                         & 4.244 & 3.975 & 3.979 & 4.833 \\
6 & 2 & $2\rm{p}^2\ ^3\rm{P}_{1}$                         & 0.016 & 0.014 & 0.014 & 0.014 & 8 & 6 & $[1\rm{s}]2\rm{p}^5\ ^3\rm{P}_{2}^{\rm o}$        & 531.4 & 528.8 & 528.2 & 529.7 \\
6 & 3 & $2\rm{p}^2\ ^3\rm{P}_{2}$                         & 0.044 & 0.037 & 0.040 & 0.040 & 8 & 7 & $[1\rm{s}]2\rm{p}^5\ ^3\rm{P}_{1}^{\rm o}$        & 531.4 & 528.8 & 528.2 & 529.7 \\
6 & 4 & $2\rm{p}^2\ ^1\rm{D}_{2}$                         & 2.958 & 2.750 & 2.848 & 2.711 & 8 & 8 & $[1\rm{s}]2\rm{p}^5\ ^3\rm{P}_{0}^{\rm o}$        & 531.4 & 528.9 & 528.3 & 529.7 \\
6 & 5 & $2\rm{p}^2\ ^1\rm{S}_{0}$                         & 5.390 & 5.000 & 5.237 & 6.089 & 8 & 9 & $[1\rm{s}]2\rm{p}^5\ ^1\rm{P}_{1}^{\rm o}$        & 533.5 & 531.2 & 530.8 & 532.7 \\
6 & 6 & $[1\rm{s}]2\rm{p}^3\ ^3\rm{D}_{3}^{\rm o}$        & 537.5 & 536.9 & 537.0 & 536.9 &   &   &                                                   &       &       &       &       \\
6 & 7 & $[1\rm{s}]2\rm{p}^3\ ^3\rm{D}_{2}^{\rm o}$        & 537.5 & 536.9 & 537.0 & 536.9 &   &   &                                                   &       &       &       &       \\
6 & 8 & $[1\rm{s}]2\rm{p}^3\ ^3\rm{D}_{1}^{\rm o}$        & 537.5 & 536.9 & 537.0 & 537.0 &   &   &                                                   &       &       &       &       \\
6 & 9 & $[1\rm{s}]2\rm{p}^3\ ^3\rm{S}_{1}^{\rm o}$        & 538.0 & 537.4 & 537.7 & 538.0 &   &   &                                                   &       &       &       &       \\
6 &10 & $[1\rm{s}]2\rm{p}^3\ ^3\rm{P}_{2}^{\rm o}$        & 538.9 & 538.4 & 538.6 & 539.4 &   &   &                                                   &       &       &       &       \\
6 &11 & $[1\rm{s}]2\rm{p}^3\ ^3\rm{P}_{1}^{\rm o}$        & 538.9 & 538.5 & 538.7 & 539.4 &   &   &                                                   &       &       &       &       \\
6 &12 & $[1\rm{s}]2\rm{p}^3\ ^3\rm{P}_{0}^{\rm o}$        & 538.9 & 538.5 & 538.7 & 539.4 &   &   &                                                   &       &       &       &       \\
6 &13 & $[1\rm{s}]2\rm{p}^3\ ^1\rm{D}_{2}^{\rm o}$        & 540.8 & 540.3 & 540.6 & 540.7 &   &   &                                                   &       &       &       &       \\
6 &14 & $[1\rm{s}]2\rm{p}^3\ ^1\rm{P}_{1}^{\rm o}$        & 542.2 & 541.9 & 542.2 & 543.1 &   &   &                                                   &       &       &       &       \\
\enddata
\end{deluxetable}

\begin{deluxetable}{rlrrrrr}
\tablecolumns{7} \tabletypesize{\scriptsize} \tablewidth{0pc}
\tablecaption{Comparison of valence level energies (eV)}
\tablehead{\colhead{$N$} & \colhead{Level} &\colhead{Expt$^{\rm a}$} & \colhead{AS1} &\colhead{AS2}&
\colhead{HF1}& \colhead{RM1} \\
} \startdata
5 & $2\rm{p}\ ^2\rm{P}_{1/2}^{\rm o}$                 & 0.000 & 0.000 & 0.000 & 0.000 & 0.000 \\
5 & $2\rm{p}\ ^2\rm{P}_{3/2}^{\rm o}$                 & 0.048 & 0.057 & 0.049 & 0.047 & 0.047 \\
6 & $2\rm{p}^2\ ^3\rm{P}_{0}$                         & 0.000 & 0.000 & 0.000 & 0.000 & 0.000 \\
6 & $2\rm{p}^2\ ^3\rm{P}_{1}$                         & 0.014 & 0.016 & 0.014 & 0.014 & 0.014 \\
6 & $2\rm{p}^2\ ^3\rm{P}_{2}$                         & 0.038 & 0.044 & 0.037 & 0.040 & 0.040 \\
6 & $2\rm{p}^2\ ^1\rm{D}_{2}$                         & 2.514 & 2.958 & 2.750 & 2.848 & 2.711 \\
6 & $2\rm{p}^2\ ^1\rm{S}_{0}$                         & 5.354 & 5.390 & 5.000 & 5.237 & 6.089 \\
7 & $2\rm{p}^3\ ^4\rm{S}_{3/2}^{\rm o}$               & 0.000 & 0.000 & 0.000 & 0.000 & 0.000 \\
7 & $2\rm{p}^3\ ^2\rm{D}_{5/2}^{\rm o}$               & 3.324 & 3.972 & 3.711 & 3.770 & 3.615 \\
7 & $2\rm{p}^3\ ^2\rm{D}_{3/2}^{\rm o}$               & 3.327 & 3.975 & 3.714 & 3.770 & 3.614 \\
7 & $2\rm{p}^3\ ^2\rm{P}_{3/2}^{\rm o}$               & 5.017 & 5.393 & 5.064 & 5.131 & 5.655 \\
7 & $2\rm{p}^3\ ^2\rm{P}_{1/2}^{\rm o}$               & 5.018 & 5.395 & 5.066 & 5.131 & 5.654 \\
8 & $2\rm{p}^4\ ^3\rm{P}_{2}$                         & 0.000 & 0.000 & 0.000 & 0.000 & 0.000 \\
8 & $2\rm{p}^4\ ^3\rm{P}_{1}$                         & 0.020 & 0.023 & 0.020 & 0.018 & 0.000 \\
8 & $2\rm{p}^4\ ^3\rm{P}_{0}$                         & 0.028 & 0.033 & 0.029 & 0.027 & 0.000 \\
8 & $2\rm{p}^4\ ^1\rm{D}_{2}$                         & 1.967 & 2.353 & 2.181 & 2.200 & 2.149 \\
8 & $2\rm{p}^4\ ^1\rm{S}_{0}$                         & 4.190 & 4.244 & 3.975 & 3.979 & 4.833 \\
\enddata
\tablenotetext{a}{Spectroscopic tables \citep{moo98}}
\end{deluxetable}
\begin{deluxetable}{rlrrll}
\tablecolumns{6} \tabletypesize{\scriptsize} \tablewidth{0pc}
\tablecaption{Comparison of K-vacancy level energies$^*$ (eV)}
\tablehead{ \colhead{N} & \colhead{State} & \colhead{AS2} & \colhead{HF1} & \colhead{Expt}&\colhead{Theory}\\
} \startdata
2 & $[1\rm{s}]2\rm{p}\ ^3\rm{P}^{\rm o}_{1}$                       & 567.7 & 568.1 & 568.6$^{\rm a}$                                               & 568.2$^{\rm e}$ \\
2 & $[1\rm{s}]2\rm{p}\ ^1\rm{P}^{\rm o}_{1}$                       & 573.5 & 573.7 & 574.0$^{\rm a}$                                               & 573.6$^{\rm e}$ \\
3 & $[1\rm{s}]2\rm{s}2\rm{p}(^3{\rm P}^{\rm o})\ ^2\rm{P}^{\rm o}$ & 563.1 & 562.7 & 562.6$^{\rm a}$                                               & 562.3$^{\rm f}$ \\
3 & $[1\rm{s}]2\rm{s}2\rm{p}(^1{\rm P}^{\rm o})\ ^2\rm{P}^{\rm o}$ & 568.6 & 568.1 & 568.2$^{\rm a}$                                               & 567.0$^{\rm f}$ \\
4 & $[1\rm{s}]2\rm{p}\ ^1\rm{P}_{1}^{\rm o}$                       & 554.5 & 554.4 &                                                               & 554.6$^{\rm f}$, 553.15$^{\rm g}$ \\
5 & $[1\rm{s}]2\rm{p}^2\ ^2\rm{D}_{3/2}$                           & 544.6 & 544.3 &                                                               & 545.8$^{\rm f}$ \\
5 & $[1\rm{s}]2\rm{p}^2\ ^2\rm{P}$                                 & 545.3 & 545.0 &                                                               & 547.0$^{\rm f}$ \\
5 & $[1\rm{s}]2\rm{p}^2\ ^2\rm{S}_{1/2}$                           & 547.1 & 547.3 &                                                               & 552.1$^{\rm f}$ \\
6 & $[1\rm{s}]2\rm{p}^3\ ^3\rm{D}_{1}^{\rm o}$                     & 536.9 & 537.0 &                                                               & 537.2$^{\rm f}$ \\
6 & $[1\rm{s}]2\rm{p}^3\ ^3\rm{S}_{1}^{\rm o}$                     & 537.4 & 537.7 &                                                               & 538.6$^{\rm f}$ \\
6 & $[1\rm{s}]2\rm{p}^3\ ^3\rm{P}_{1}^{\rm o}$                     & 538.5 & 538.7 &                                                               & 540.7$^{\rm f}$ \\
7 & $[1\rm{s}]2\rm{p}^4\ ^4\rm{P}$                                 & 531.9 & 531.7 & 530.8(3)$^{\rm b}$, 530.41(4)$^{\rm c}$                       & 532.9$^{\rm f}$, 531.0$^{\rm h}$, 534.1$^{\rm i}$ \\
7 & $[1\rm{s}]2\rm{p}^4\ ^2\rm{D}$                                 & 536.1 & 536.1 &                                                               & 535.3$^{\rm h}$, 538.7$^{\rm i}$ \\
7 & $[1\rm{s}]2\rm{p}^4\ ^2\rm{P}$                                 & 536.2 & 536.3 & 535.7(3)$^{\rm b}$, 535.23(4)$^{\rm c}$                       & 535.9$^{\rm h}$, 539.2$^{\rm i}$ \\
7 & $[1\rm{s}]2\rm{p}^4\ ^2\rm{S}$                                 & 538.4 & 538.3 &                                                               & 538.1$^{\rm h}$, 542.9$^{\rm i}$ \\
8 & $[1\rm{s}]2\rm{p}^5\ ^3\rm{P}^{\rm o}$                         & 528.8 & 528.2 & 527.2(3)$^{\rm b}$, 526.79(4)$^{\rm c}$, 527.85(10)$^{\rm d}$ & 528.8$^{\rm f}$, 528.22$^{\rm i}$, 528.33$^{\rm i}$ \\
\enddata
\tablenotetext{*}{Relative to each ion ground level}
\tablenotetext{a}{Spectroscopic tables \citep{moo98}}
\tablenotetext{b}{Auger electron spectrometry \citep{kra94,cal94}}
\tablenotetext{c}{Photoionization experiment \citep{sto97}}
\tablenotetext{d}{Photoabsorption measurements \citep{men96}}
\tablenotetext{e}{Non-relativistic CI calculation \citep{cha00}}
\tablenotetext{f}{$R$-matrix calculation by \citet{pra03}}
\tablenotetext{g}{Multiconfiguration Dirac--Fock calculation by \citet{che85}}
\tablenotetext{h}{$R$-matrix calculation \citep{gor00b}}
\tablenotetext{i}{$R$-matrix calculation \citep{mcl98}}
\end{deluxetable}


\begin{deluxetable}{rrrrrrrrrrrrrrrrr}
\tablecolumns{17} \tabletypesize{\scriptsize} \tablewidth{0pc}
\tablecaption{Calculated wavelengths and $gf$-values for K$\alpha$
transitions}
\tablehead{ \colhead{}&\colhead{}&\colhead{}&
\multicolumn{2}{c}{AS2}&&\multicolumn{2}{c}{HF1} &
\colhead{}&\colhead{}&\colhead{}&\colhead{}&
\multicolumn{2}{c}{AS2}&&\multicolumn{2}{c}{HF1}\\
\cline{4-5} \cline{7-8} \cline{13-14} \cline{16-17} \\
\colhead{N} & \colhead{$i$} & \colhead{$j$} & \colhead{$\lambda$} &
\colhead{$gf_{ij}$} && \colhead{$\lambda$}& \colhead{$gf_{ij}$} &&
\colhead{N} & \colhead{$i$} & \colhead{$j$} & \colhead{$\lambda$} &
\colhead{$gf_{ij}$} && \colhead{$\lambda$}& \colhead{$gf_{ij}$} \\
\colhead{} & \colhead{} & \colhead{} & \colhead{(\AA)} & \colhead{} &&
\colhead{(\AA)} & \colhead{} && \colhead{} & \colhead{} & \colhead{} &
\colhead{(\AA)} & \colhead{} && \colhead{(\AA)} & \colhead{} \\
}
\startdata
 2 & 1 &  2 &  21.84 & 1.22E$-$4 && 21.82 & 7.35E$-$5 && 6 & 4 & 13 &  23.06 & 1.35E$+$0 && 23.06 & 1.35E$+$0 \\
 2 & 1 &  3 &  21.62 & 8.27E$-$1 && 21.61 & 7.96E$-$1 && 6 & 4 & 14 &  23.00 & 5.09E$-$1 && 22.99 & 4.98E$-$1 \\
 3 & 1 &  2 &  22.02 & 4.93E$-$1 && 22.04 & 4.37E$-$1 && 6 & 5 & 14 &  23.09 & 3.37E$-$1 && 23.09 & 3.41E$-$1 \\
 3 & 1 &  3 &  22.02 & 9.95E$-$1 && 22.03 & 8.81E$-$1 && 7 & 1 &  6 &  23.31 & 4.24E$-$1 && 23.32 & 4.56E$-$1 \\
 3 & 1 &  4 &  21.80 & 4.53E$-$2 && 21.83 & 4.40E$-$2 && 7 & 1 &  7 &  23.31 & 2.83E$-$1 && 23.32 & 3.04E$-$1 \\
 3 & 1 &  5 &  21.80 & 8.23E$-$2 && 21.83 & 8.09E$-$2 && 7 & 1 &  8 &  23.31 & 1.41E$-$1 && 23.32 & 1.52E$-$1 \\
 4 & 1 &  2 &  22.36 & 7.08E$-$1 && 22.37 & 6.52E$-$1 && 7 & 2 &  9 &  23.29 & 6.22E$-$1 && 23.29 & 6.70E$-$1 \\
 5 & 1 &  4 &  22.77 & 3.53E$-$1 && 22.78 & 3.33E$-$1 && 7 & 2 & 10 &  23.29 & 1.00E$-$1 && 23.29 & 1.45E$-$2 \\
 5 & 1 &  5 &  22.74 & 4.36E$-$1 && 22.75 & 4.09E$-$1 && 7 & 2 & 11 &  23.29 & 6.72E$-$1 && 23.28 & 8.18E$-$1 \\
 5 & 1 &  6 &  22.74 & 1.89E$-$1 && 22.75 & 1.80E$-$1 && 7 & 3 &  9 &  23.29 & 4.09E$-$2 && 23.29 & 4.43E$-$2 \\
 5 & 1 &  7 &  22.66 & 7.15E$-$2 && 22.65 & 6.67E$-$2 && 7 & 3 & 10 &  23.29 & 4.76E$-$1 && 23.29 & 4.74E$-$1 \\
 5 & 2 &  3 &  22.77 & 5.90E$-$1 && 22.78 & 5.61E$-$1 && 7 & 3 & 11 &  23.29 & 1.48E$-$2 && 23.28 & 5.42E$-$2 \\
 5 & 2 &  4 &  22.77 & 4.21E$-$2 && 22.78 & 4.24E$-$2 && 7 & 3 & 12 &  23.29 & 3.98E$-$1 && 23.28 & 4.30E$-$1 \\
 5 & 2 &  5 &  22.74 & 2.06E$-$1 && 22.75 & 1.95E$-$1 && 7 & 4 &  9 &  23.35 & 2.00E$-$1 && 23.35 & 2.19E$-$1 \\
 5 & 2 &  6 &  22.74 & 1.09E$+$0 && 22.75 & 1.03E$+$0 && 7 & 4 & 10 &  23.35 & 2.04E$-$1 && 23.35 & 4.41E$-$2 \\
 5 & 2 &  7 &  22.66 & 1.68E$-$1 && 22.66 & 1.52E$-$1 && 7 & 4 & 11 &  23.35 & 1.66E$-$1 && 23.34 & 3.61E$-$1 \\
 6 & 1 &  8 &  23.09 & 1.41E$-$1 && 23.09 & 1.42E$-$1 && 7 & 4 & 12 &  23.34 & 7.48E$-$2 && 23.34 & 8.09E$-$2 \\
 6 & 1 &  9 &  23.07 & 1.30E$-$1 && 23.06 & 1.30E$-$1 && 7 & 4 & 13 &  23.25 & 2.12E$-$1 && 23.25 & 2.29E$-$1 \\
 6 & 1 & 11 &  23.03 & 8.40E$-$2 && 23.02 & 8.18E$-$2 && 7 & 5 & 10 &  23.35 & 1.38E$-$2 && 23.35 & 9.82E$-$2 \\
 6 & 2 &  7 &  23.09 & 3.17E$-$1 && 23.09 & 3.18E$-$1 && 7 & 5 & 11 &  23.35 & 1.67E$-$1 && 23.34 & 1.00E$-$1 \\
 6 & 2 &  8 &  23.09 & 1.00E$-$1 && 23.09 & 9.95E$-$2 && 7 & 5 & 12 &  23.34 & 1.46E$-$1 && 23.34 & 1.61E$-$1 \\
 6 & 2 &  9 &  23.07 & 3.82E$-$1 && 23.06 & 3.82E$-$1 && 7 & 5 & 13 &  23.25 & 1.01E$-$1 && 23.25 & 1.08E$-$1 \\
 6 & 2 & 10 &  23.03 & 1.07E$-$1 && 23.02 & 1.05E$-$1 && 8 & 1 &  6 &  23.45 & 4.31E$-$1 && 23.47 & 5.25E$-$1 \\
 6 & 2 & 11 &  23.03 & 6.52E$-$2 && 23.02 & 6.31E$-$2 && 8 & 1 &  7 &  23.44 & 1.44E$-$1 && 23.47 & 1.75E$-$1 \\
 6 & 2 & 12 &  23.03 & 9.31E$-$2 && 23.02 & 9.12E$-$2 && 8 & 2 &  6 &  23.45 & 1.44E$-$1 && 23.47 & 1.80E$-$1 \\
 6 & 3 &  6 &  23.10 & 5.75E$-$1 && 23.09 & 5.77E$-$1 && 8 & 2 &  7 &  23.45 & 8.61E$-$2 && 23.47 & 1.08E$-$1 \\
 6 & 3 &  7 &  23.09 & 9.36E$-$2 && 23.09 & 9.44E$-$2 && 8 & 2 &  8 &  23.44 & 1.15E$-$1 && 23.47 & 1.44E$-$1 \\
 6 & 3 &  8 &  23.09 & 5.77E$-$3 && 23.09 & 6.05E$-$3 && 8 & 3 &  7 &  23.45 & 1.15E$-$1 && 23.47 & 1.44E$-$1 \\
 6 & 3 &  9 &  23.07 & 6.11E$-$1 && 23.06 & 6.08E$-$1 && 8 & 4 &  9 &  23.44 & 6.28E$-$1 && 23.45 & 7.85E$-$1 \\
 6 & 3 & 10 &  23.03 & 3.58E$-$1 && 23.02 & 3.51E$-$1 && 8 & 5 &  9 &  23.52 & 1.05E$-$1 && 23.53 & 1.32E$-$1 \\
 6 & 3 & 11 &  23.03 & 1.31E$-$1 && 23.02 & 1.29E$-$1 &&   &   &    &        &           &&       &           \\
\enddata
\end{deluxetable}


\begin{deluxetable}{rllrrll}
\tablecolumns{6} \tabletypesize{\scriptsize} \tablewidth{0pc}
\tablecaption{Wavelength (\AA) comparison for K$\alpha$
transitions} \tablehead{ \colhead{N} & \colhead{$i$} &
\colhead{$j$} & \colhead{AS2} & \colhead{HF1} &
\colhead{Expt}&\colhead{Other theory}\\
}
\startdata
2 & $1\rm{s}^2\ ^1\rm{S}_0$           & $[1\rm{s}]2\rm{p}\ ^3\rm{P}^{\rm o}_1$                         & 21.84 & 21.82 & 21.807$^{\rm a}$                                            & 21.84$^{\rm i}$, 21.84$^{\rm j}$\\
  & $1\rm{s}^2\ ^1\rm{S}_0$           & $[1\rm{s}]2\rm{p}\ ^1\rm{P}^{\rm o}_1$                         & 21.62 & 21.61 & 21.6020(3)$^{\rm b}$                                        & 21.60$^{\rm i}$\\
3 & $2\rm{s}\ ^2\rm{S}  $             & $[1\rm{s}]2\rm{s}2\rm{p}(^3{\rm P}^{\rm o})\ ^2\rm{P}^{\rm o}$ & 22.02 & 22.03 & 22.038$^{\rm a}$, 22.01(1)$^{\rm c}$, 22.0194(16)$^{\rm d}$ & 22.02$^{\rm i}$, 22.06$^{\rm j}$, 22.05$^{\rm k}$, 22.00$^{\rm l}$\\
  & $2\rm{s}\ ^2\rm{S}  $             & $[1\rm{s}]2\rm{s}2\rm{p}(^1{\rm P}^{\rm o})\ ^2\rm{P}^{\rm o}$ & 21.80 & 21.83 & 21.82$^{\rm a}$                                             & 21.84$^{\rm j}$,21.87$^{\rm k}$, 21.79$^{\rm l}$\\
4 & $2\rm{s}^2\ ^1\rm{S}$             & $[1\rm{s}]2\rm{p}\ ^1\rm{P}^{\rm o}$                           & 22.36 & 22.37 & 22.38(1)$^{\rm c}$, 22.374(3)$^{\rm d}$                     & 22.38$^{\rm i}$, 22.41$^{\rm j}$, 22.35$^{\rm k}$, 22.33$^{\rm l}$\\
5 & $2\rm{p}\ ^2\rm{P}_{1/2}^{\rm o}$ & $[1\rm{s}]2\rm{p}^2\ ^2\rm{D}_{3/2}$                           & 22.77 & 22.78 & 22.74(2)$^{\rm c}$                                          & 22.73$^{\rm k}$, 22.73$^{\rm l}$\\
  & $2\rm{p}\ ^2\rm{P}_{1/2}^{\rm o}$ & $[1\rm{s}]2\rm{p}^2\ ^2\rm{P}$                                 & 22.74 & 22.75 &                                                             & 22.67$^{\rm k}$, 22.78$^{\rm l}$\\
  & $2\rm{p}\ ^2\rm{P}_{1/2}^{\rm o}$ & $[1\rm{s}]2\rm{p}^2\ ^2\rm{S}_{1/2}$                           & 22.66 & 22.65 &                                                             & 22.46$^{\rm k}$, 22.73$^{\rm l}$\\
6 & $2\rm{p}^2\ ^3\rm{P}_{0}$         & $[1\rm{s}]2\rm{p}^3\ ^3\rm{D}_{1}^{\rm o}$                     & 23.09 & 23.09 & 23.17(1)$^{\rm c}$                                          & 23.08$^{\rm k}$, 23.11$^{\rm l}$\\
  & $2\rm{p}^2\ ^3\rm{P}_{0}$         & $[1\rm{s}]2\rm{p}^3\ ^3\rm{S}_{1}^{\rm o}$                     & 23.07 & 23.06 & 23.00(2)$^{\rm c}$                                          & 23.02$^{\rm k}$, 23.05$^{\rm l}$\\
  & $2\rm{p}^2\ ^3\rm{P}_{0}$         & $[1\rm{s}]2\rm{p}^3\ ^3\rm{P}_{1}^{\rm o}$                     & 23.03 & 23.02 &                                                             & 22.93$^{\rm k}$, 22.98$^{\rm l}$\\
7 & $2\rm{p}^3\ ^4\rm{S}^{\rm o}$     & $[1\rm{s}]2\rm{p}^4\ ^4\rm{P}$                                 & 23.31 & 23.32 & 23.36(1)$^{\rm e}$                                          & 23.27$^{\rm k}$, 23.30$^{\rm l}$\\
  & $2\rm{p}^3\ ^2\rm{D}^{\rm o}$     & $[1\rm{s}]2\rm{p}^4\ ^2\rm{P}$                                 & 23.29 & 23.28 & 23.29(1)$^{\rm e}$                                          &                \\
  & $2\rm{p}^3\ ^2\rm{P}^{\rm o}$     & $[1\rm{s}]2\rm{p}^4\ ^2\rm{P}$                                 & 23.35 & 23.34 & 23.36(1)$^{\rm e}$                                          &                \\
8 & $2\rm{p}^4\ ^3\rm{P}$             & $[1\rm{s}]2\rm{p}^5\ ^3\rm{P}^{\rm o}$                         & 23.45 & 23.47 & 23.52(3)$^{\rm e}$, 23.489(5)$^{\rm f}$, 23.536(2)$^{\rm g}$& 23.45$^{\rm k}$\\
  &                                   &                                                                &       &       & 23.508(3)$^{\rm h}$                                         &                \\
\enddata
\tablenotetext{a}{Spectroscopic tables \citep{moo98}}
\tablenotetext{b}{Spectroscopic measurements \citep{eng95}}
\tablenotetext{c}{Spectroscopy of NGC 5548 \citep{ste03}}
\tablenotetext{d}{Electron beam ion trap measurements \citep{sch04}}
\tablenotetext{e}{Auger electron spectrometry \citep{kra94,cal94}}
\tablenotetext{f}{Photoabsorption measurements \citep{men96}}
\tablenotetext{g}{Photoionization experiment \citep{sto97}}
\tablenotetext{h}{ISM observations \citep{jue04}}
\tablenotetext{i}{{\sc 1/Z} expansion method \citep{vai71,vai78}}
\tablenotetext{j}{Multiconfiguration Dirac--Fock calculation \citep{che85,che86}}
\tablenotetext{k}{$R$-matrix calculation \citep{pra03}}
\tablenotetext{l}{{\sc hullac} calculation \citep{beh02}}
\end{deluxetable}


\begin{deluxetable}{rllrrr}
\tablecolumns{6} \tabletypesize{\scriptsize} \tablewidth{0pc}
\tablecaption{Comparison of theoretical $f$-values for K$\alpha$
transitions} \tablehead{ \colhead{N} & \colhead{$i$} &
\colhead{$j$} & \colhead{AS2} & \colhead{HF1} &
\colhead{RM2$^{\rm a}$}\\
}
\startdata
 3 & ${\rm 2s}\ ^2{\rm S}$                 & $[{\rm 1s}]{\rm 2s2p}(^3{\rm P}^{\rm o})\ ^2{\rm P}^{\rm o}$ & 0.744 & 0.659 & 0.576 \\
 3 & ${\rm 2s}\ ^2{\rm S}$                 & $[{\rm 1s}]{\rm 2s2p}(^1{\rm P}^{\rm o})\ ^2{\rm P}^{\rm o}$ & 0.077 & 0.062 & 0.061 \\
 4 & ${\rm 2s}^2\ ^1{\rm S}$               & $[{\rm 1s}]{\rm 2p}\ ^1{\rm P}^{\rm o}$                      & 0.708 & 0.652 & 0.565 \\
 5 & ${\rm 2p}\ ^2{\rm P}^{\rm o}_{1/2}$   & $[{\rm 1s}]{\rm 2p}^2\ ^2{\rm D}_{3/2}$                      & 0.177 & 0.167 & 0.132 \\
 5 & ${\rm 2p}\ ^2{\rm P}^{\rm o}_{1/2}$   & $[{\rm 1s}]{\rm 2p}^2\ ^2{\rm P}$                            & 0.313 & 0.295 & 0.252 \\
 5 & ${\rm 2p}\ ^2{\rm P}^{\rm o}_{1/2}$   & $[{\rm 1s}]{\rm 2p}^2\ ^2{\rm S}_{1/2}$                      & 0.036 & 0.033 & 0.027 \\
 6 & ${\rm 2p}^2\ ^3{\rm P}_0$             & $[{\rm 1s}]{\rm 2p}^3\ ^3{\rm D}^{\rm o}_1$                  & 0.141 & 0.142 & 0.119 \\
 6 & ${\rm 2p}^2\ ^3{\rm P}_0$             & $[{\rm 1s}]{\rm 2p}^3\ ^3{\rm S}^{\rm o}_1$                  & 0.130 & 0.130 & 0.102 \\
 6 & ${\rm 2p}^2\ ^3{\rm P}_0$             & $[{\rm 1s}]{\rm 2p}^3\ ^3{\rm P}^{\rm o}_1$                  & 0.084 & 0.082 & 0.067 \\
 7 & ${\rm 2p}^3\ ^4{\rm S}^{\rm o}_{3/2}$ & $[{\rm 1s}]{\rm 2p}^4\ ^4{\rm P}$                            & 0.212 & 0.228 & 0.184 \\
 8 & ${\rm 2p}^4\ ^3{\rm P}_2$             & $[{\rm 1s}]{\rm 2p}^5\ ^3{\rm P}$                            & 0.115 & 0.140 & 0.113 \\
\enddata
\tablenotetext{a}{$R$-matrix data by \citet{pra03}}
\end{deluxetable}


\begin{deluxetable}{rllrrrlrr}
\tablecolumns{9} \tabletypesize{\scriptsize} \tablewidth{0pc}
\tablecaption{Comparison of branching ratios for KLL Auger
transitions} \tablehead{ \colhead{N} & \colhead{$i$} &
\colhead{$j$} & \colhead{AS1} & \colhead{AS2} & \colhead{HF1} &
\colhead{Expt$^{\rm a}$} & \colhead{MCDF$^{\rm b}$} &
\colhead{CC$^{\rm c}$}\\
} \startdata
7 & $[1\rm{s}]2\rm{p}^4\ ^4\rm{P}$         & $2\rm{p}^2$            & 0.42 & 0.42 & 0.43 & 0.55(13)& 0.36 & 0.43 \\
  &                                        & $[2\rm{s}]2\rm{p}^3$   & 0.41 & 0.40 & 0.40 & 0.33(3) & 0.46 & 0.42 \\
  &                                        & $[2\rm{s}^2]2\rm{p}^4$ & 0.17 & 0.18 & 0.18 & 0.12(2) & 0.18 & 0.15 \\
7 & $[1\rm{s}]2\rm{p}^4\ ^2\rm{P}$         & $2\rm{p}^2$            & 0.53 & 0.53 & 0.51 & 0.53(9) & 0.45 & 0.47 \\
  &                                        & $[2\rm{s}]2\rm{p}^3$   & 0.27 & 0.27 & 0.28 & 0.28(4) & 0.31 & 0.28 \\
  &                                        & $[2\rm{s}^2]2\rm{p}^4$ & 0.20 & 0.20 & 0.21 & 0.19(3) & 0.24 & 0.25 \\
8 & $[1\rm{s}]2\rm{p}^5\ ^3\rm{P}^{\rm o}$ & $2\rm{p}^3$            & 0.53 & 0.51 & 0.55 & 0.606   &      &      \\
  &                                        & $[2\rm{s}]2\rm{p}^4$   & 0.35 & 0.35 & 0.32 & 0.30    &      &      \\
  &                                        & $[2\rm{s}^2]2\rm{p}^5$ & 0.12 & 0.14 & 0.13 & 0.094   &      &      \\
\enddata
\tablenotetext{a}{Auger electron spectrometry \citep{cal93,cal94}}
\tablenotetext{b}{MCDF calculation by M. H. Chen as quoted in \citet{cal94}}
\tablenotetext{c}{Close-coupling calculation \citep{pet94}}
\end{deluxetable}


\begin{deluxetable}{rrlrrrrr}
\tablecolumns{8} \tabletypesize{\scriptsize} \tablewidth{0pc}
\tablecaption{Calculated radiative and Auger widths for K-vacancy levels}
\tablehead{
\colhead{}&\colhead{}&\colhead{}& \multicolumn{2}{c}{AS2} && \multicolumn{2}{c}{HF1}\\
\cline{4-5} \cline{7-8}\\
\colhead{N} & \colhead{$i$} & \colhead{Level} & \colhead{$A_{\rm r}$} &
\colhead{$A_{\rm a}$}& & \colhead{$A_{\rm r}$} & \colhead{$A_{\rm a}$}\\
\colhead{} & \colhead{} & \colhead{} & \colhead{(s$^{-1}$)} &
\colhead{(s$^{-1}$)}& & \colhead{(s$^{-1}$)} & \colhead{(s$^{-1}$)}\\
} \startdata
2 & 2 & $[1\rm{s}]2\rm{p}\ ^3\rm{P}^{\rm o}_{1}$          & 6.60E+08 &           &&  3.44E+08 &          \\
2 & 3 & $[1\rm{s}]2\rm{p}\ ^1\rm{P}^{\rm o}_{1}$          & 3.93E+12 &           &&  3.80E+12 &          \\
3 & 2 & $[1\rm{s}]2\rm{s}2\rm{p}\ ^2\rm{P}_{1/2}^{\rm o}$ & 3.39E+12 &  6.60E+12 &&  3.00E+12 & 7.53E+12 \\
3 & 3 & $[1\rm{s}]2\rm{s}2\rm{p}\ ^2\rm{P}_{3/2}^{\rm o}$ & 3.42E+12 &  5.76E+12 &&  3.02E+12 & 6.96E+12 \\
3 & 4 & $[1\rm{s}]2\rm{s}2\rm{p}\ ^2\rm{P}_{1/2}^{\rm o}$ & 3.19E+11 &  7.56E+13 &&  3.12E+11 & 6.67E+13 \\
3 & 5 & $[1\rm{s}]2\rm{s}2\rm{p}\ ^2\rm{P}_{3/2}^{\rm o}$ & 2.90E+11 &  7.60E+13 &&  2.87E+11 & 6.73E+13 \\
4 & 2 & $[1\rm{s}]2\rm{p}\ ^1\rm{P}_{1}^{\rm o}$          & 3.40E+12 &  9.10E+13 &&  3.10E+12 & 8.15E+13 \\
5 & 3 & $[1\rm{s}]2\rm{p}^2\ ^2\rm{D}_{5/2}$              & 1.34E+12 &  2.19E+14 &&  1.22E+12 & 1.93E+14 \\
5 & 4 & $[1\rm{s}]2\rm{p}^2\ ^2\rm{D}_{3/2}$              & 1.35E+12 &  2.18E+14 &&  1.28E+12 & 1.92E+14 \\
5 & 5 & $[1\rm{s}]2\rm{p}^2\ ^2\rm{P}_{1/2}$              & 4.31E+12 &  9.29E+13 &&  4.05E+12 & 7.57E+13 \\
5 & 6 & $[1\rm{s}]2\rm{p}^2\ ^2\rm{P}_{3/2}$              & 4.31E+12 &  9.30E+13 &&  4.05E+12 & 7.58E+13 \\
5 & 7 & $[1\rm{s}]2\rm{p}^2\ ^2\rm{S}_{1/2}$              & 1.63E+12 &  1.89E+14 &&  1.50E+12 & 1.66E+14 \\
6 & 6 & $[1\rm{s}]2\rm{p}^3\ ^3\rm{D}_{3}^{\rm o}$        & 1.05E+12 &  2.37E+14 &&  1.06E+12 & 2.11E+14 \\
6 & 7 & $[1\rm{s}]2\rm{p}^3\ ^3\rm{D}_{2}^{\rm o}$        & 1.05E+12 &  2.37E+14 &&  1.06E+12 & 2.10E+14 \\
6 & 8 & $[1\rm{s}]2\rm{p}^3\ ^3\rm{D}_{1}^{\rm o}$        & 1.05E+12 &  2.37E+14 &&  1.06E+12 & 2.10E+14 \\
6 & 9 & $[1\rm{s}]2\rm{p}^3\ ^3\rm{S}_{1}^{\rm o}$        & 4.81E+12 &  8.87E+13 &&  4.81E+12 & 7.24E+13 \\
6 &10 & $[1\rm{s}]2\rm{p}^3\ ^3\rm{P}_{2}^{\rm o}$        & 1.20E+12 &  2.22E+14 &&  1.18E+12 & 1.98E+14 \\
6 &11 & $[1\rm{s}]2\rm{p}^3\ ^3\rm{P}_{1}^{\rm o}$        & 1.20E+12 &  2.22E+14 &&  1.18E+12 & 1.98E+14 \\
6 &12 & $[1\rm{s}]2\rm{p}^3\ ^3\rm{P}_{0}^{\rm o}$        & 1.20E+12 &  2.22E+14 &&  1.18E+12 & 1.98E+14 \\
6 &13 & $[1\rm{s}]2\rm{p}^3\ ^1\rm{D}_{2}^{\rm o}$        & 3.46E+12 &  2.05E+14 &&  3.46E+12 & 1.84E+14 \\
6 &14 & $[1\rm{s}]2\rm{p}^3\ ^1\rm{P}_{1}^{\rm o}$        & 3.62E+12 &  1.90E+14 &&  3.60E+12 & 1.71E+14 \\
7 & 6 & $[1\rm{s}]2\rm{p}^4\ ^4\rm{P}_{5/2}$              & 8.67E+11 &  2.25E+14 &&  9.35E+11 & 2.04E+14 \\
7 & 7 & $[1\rm{s}]2\rm{p}^4\ ^4\rm{P}_{3/2}$              & 8.67E+11 &  2.25E+14 &&  9.35E+11 & 2.04E+14 \\
7 & 8 & $[1\rm{s}]2\rm{p}^4\ ^4\rm{P}_{1/2}$              & 8.68E+11 &  2.25E+14 &&  9.35E+11 & 2.04E+14 \\
7 & 9 & $[1\rm{s}]2\rm{p}^4\ ^2\rm{D}_{5/2}$              & 1.78E+12 &  2.60E+14 &&  1.93E+12 & 2.44E+14 \\
7 &10 & $[1\rm{s}]2\rm{p}^4\ ^2\rm{D}_{3/2}$              & 2.46E+12 &  2.36E+14 &&  1.96E+12 & 2.43E+14 \\
7 &11 & $[1\rm{s}]2\rm{p}^4\ ^2\rm{P}_{3/2}$              & 3.17E+12 &  2.10E+14 &&  4.17E+12 & 1.73E+14 \\
7 &12 & $[1\rm{s}]2\rm{p}^4\ ^2\rm{P}_{1/2}$              & 3.85E+12 &  1.86E+14 &&  4.20E+12 & 1.72E+14 \\
7 &13 & $[1\rm{s}]2\rm{p}^4\ ^2\rm{S}_{1/2}$              & 1.96E+12 &  2.43E+14 &&  2.10E+12 & 2.28E+14 \\
8 & 6 & $[1\rm{s}]2\rm{p}^5\ ^3\rm{P}_{2}^{\rm o}$        & 1.39E+12 &  2.41E+14 &&  1.70E+12 & 2.48E+14 \\
8 & 7 & $[1\rm{s}]2\rm{p}^5\ ^3\rm{P}_{1}^{\rm o}$        & 1.39E+12 &  2.41E+14 &&  1.70E+12 & 2.48E+14 \\
8 & 8 & $[1\rm{s}]2\rm{p}^5\ ^3\rm{P}_{0}^{\rm o}$        & 1.39E+12 &  2.41E+14 &&  1.70E+12 & 2.48E+14 \\
8 & 9 & $[1\rm{s}]2\rm{p}^5\ ^1\rm{P}_{1}^{\rm o}$        & 2.99E+12 &  2.20E+14 &&  3.66E+12 & 2.29E+14 \\
\enddata
\end{deluxetable}


\begin{deluxetable}{lllll}
\tablecolumns{59} \tabletypesize{\scriptsize} \tablewidth{0pc}
\tablecaption{Comparison of $\Delta E(2,3)^*$ for O ions}
\tablehead{ \colhead{$N$} & \colhead{$z$} &\colhead{Present$^{\rm a}$} & \colhead{Expt} & \colhead{Other theory} \\
} \startdata
8 & 1 & 1.06 & 1.06$^{\rm b}$, 0.99$^{\rm c}$ & 1.02$^{\rm e}$, 1.00$^{\rm f}$\\
7 & 2 & 1.93 &&\\
6 & 3 & 2.90 &&\\
5 & 4 & 3.73 &&\\
4 & 5 & 5.03 && \\
3 & 6 & 5.64 &&\\
2 & 7 & 7.00 & 7.02$^{\rm d}$& \\
1 & 8 & 8.90 & 8.90$^{\rm d}$& \\
\enddata
\tablenotetext{*}{Energy interval in Ryd between the lowest $n=2$ and $n=3$ resonances}
\tablenotetext{a}{Approximation RM1 for $N>2$ and AS1 for $N\leq 2$}
\tablenotetext{b}{Photoionization experiment \citep{sto97}}
\tablenotetext{c}{Photoabsorption measurements \citep{men96}}
\tablenotetext{d}{Spectroscopic tables \citep{moo98}}
\tablenotetext{e}{$R$-matrix calculation \citep{mcl98} with the experimental thresholds of \citet{kra94}}
\tablenotetext{f}{$R$-matrix calculation \citep{mcl98} with the experimental thresholds of \citet{sto97}}
\end{deluxetable}


\end{document}